\documentclass[aps,prl,reprint,onecolumn,10pt,notitlepage,amsmath,amssymb,nofootinbib,floatfix]{revtex4-2}

\usepackage[T1]{fontenc}
\usepackage[utf8]{inputenc}
\usepackage{amsmath,amssymb,mathrsfs,bm,mathtools,booktabs}
\usepackage{graphicx}
\usepackage[hidelinks]{hyperref}
\newcommand{\Tr}{\operatorname{Tr}}

\newcommand{\W}{\mathcal W}
\newcommand{\Op}{\mathcal O_P}

\newcommand{\ket}[1]{\lvert #1\rangle}

\newcommand{\Ec}{\mathcal E_{\mathrm C}}

\newcommand{\ep}{\mathcal E(P)}
\newcommand{\emin}{\mathcal E_{\min}}
\newcommand{\diss}{\mathscr D}
\newcommand{\dd}{\mathrm d}

\begin{document}
\title{Charge-locked ergotropy in a Josephson-coupled quantum battery
}
\author{Antti J. Niemi}
\affiliation{Nordita, Stockholm University, Hannes Alfv\'ens v\"ag 12, SE-106 91 Stockholm, Sweden}
\affiliation{Wilczek Quantum Center, Shanghai Institute for Advanced Studies, University of Science and 
Technology of China, Shanghai 201315, China}

\begin{abstract}
Charge-locked ergotropy is the work that cannot be extracted unitarily from a quantum battery without changing its 
mean total charge. We study it in a pair of Josephson-coupled superconducting islands. Adding one Cooper pair to 
each island in joint thermal equilibrium locks all the added energy, but adding two to one island locks exactly half. 
At fixed mean charge the energy-minimizing state need not be stationary, and its phase evolves at frequencies that 
Josephson coupling splits by charge parity. A prescribed-current discharge model separates charge-transfer energy
from internal loss.
\end{abstract}
\maketitle

%
%
%
%
%
%

\paragraph*{Introduction.}
Quantum batteries store energy in prepared quantum states of spin and qubit
arrays~\cite{Alicki2013,Ferraro2018,Paulino2026}.
Their charging and discharging proceed through controlled
interactions~\cite{Binder2015}.
Collective effects can enhance charging power~\cite{Campaioli2017},
and quantum coherence can increase the work available~\cite{Francica2020}.
The maximum work extractable through cyclic unitary transformations
is called ergotropy~\cite{Allahverdyan2004,Pusz1978}. However,
charge conservation can restrict these transformations,
leaving part of the ergotropy inaccessible~\cite{Mitsuhashi2022}.

Here we investigate the ergotropy that becomes locked when unitary extraction
preserves the mean total charge.
This condition is weaker than strict charge conservation and
therefore gives a lower bound on the ergotropy locked by the latter.
Motivated by superconducting quantum batteries~\cite{Hu2022},
we consider two identical Josephson-coupled charge islands.
Tunneling transfers Cooper pairs coherently between the
islands~\cite{Bouchiat1998,Nakamura1999}, preserving
their total charge, whereas releasing the locked energy
requires charge exchange with an external system.

We find that for a thermal reference state, an exact charge-shift relation
fixes the locked ergotropy at every even mean charge,
independently of temperature and Josephson coupling.
Between these values, the locked ergotropy obeys bounds
that are likewise independent of both.
At fixed density-matrix spectrum and mean charge,
an energy-minimizing state can retain coherence between
charge sectors and evolve under the time-independent
Hamiltonian.
Its phase evolution rate is set by the charge derivative
of the locked ergotropy~\cite{Niemi2026}, with Josephson
coupling splitting the principal frequencies according
to charge parity.
Adding two Cooper pairs to one island leaves exactly half
the added energy charge-locked, while the other half remains
extractable at fixed mean charge, in a quantum counterpart
of the two-capacitor problem~\cite{Zucker1955,Mita1999,Boykin2002,Hoekstra2020}.
We illustrate relaxation and prescribed-current discharge
using the symmetry-covariant generalization~\cite{Niemi2026}
of quantum Landau--Lifshitz--Gilbert (qLLG)
dynamics~\cite{Liu2024}.


%
%
%
%
%
%

\paragraph*{Charge-locked ergotropy.}
We consider a Hamiltonian $H$ with conserved charges $Q_a$, $[H,Q_a]=0$.
Let $P=\{p_k\}$ be the spectrum of an initial density matrix $\rho_0$ and
let $\Op=\{U\rho_0 U^\dagger\}$ denote its unitary orbit, the set of
isospectral density matrices. For a trajectory $\rho(t)\in \Op$ with $\rho(0)=\rho_0$, write
$E(t)=E(\rho)= \Tr[\rho (t) H] $ and $q_a(t)=q_a(\rho)= \Tr[\rho (t)\, Q_a]$.
For a prescribed admissible charge vector
$\bm q=(q_a)$, define
\begin{equation}
\ep =\min_{\rho\in\Op} E(\rho)  \ \ \ \&  \ \ \  
\emin(P, \bm q) = \!\!\!\!\!
\min_{\substack{\rho\in\Op\\
q_a(\rho)=q_a\ \forall a
}} 
\hspace{-0.3cm} E(\rho).
\label{eq:min}
\end{equation}
Since the constraints fix only 
charge expectations, rather than definite
charge eigenvalues, the admissible states may retain coherence between
different charge sectors. The ergotropy of a state $\rho\in\Op$ satisfying
$\bm q(\rho)=\bm q$ is $\W(\rho;H)=E(\rho)-\ep$~\cite{Allahverdyan2004,Pusz1978},
which we decompose as
\begin{equation}
\W(\rho;H)=
\underbrace{E(\rho)-\emin(P,\bm q)}_{\W_{\rm free}(\rho)}
+\underbrace{\emin(P,\bm q)-\ep}_{\W_Q(P,\bm q)}.
\label{eq:split}
\end{equation}
Both terms are non-negative. The first, $\W_{\rm free}(\rho)$, is the maximum
work extractable by unitaries that leave all charge expectations
unchanged between the initial and final states. The second, $\W_Q(P,\bm q)$,
is the charge-locked ergotropy, which can be extracted only by changing
the charge expectations. 
At a state \(\rho_*\) attaining \(\emin(P,\bm q)\), we have \(\W_{\rm free}(\rho_*)=0\), so 
that its entire ergotropy is charge-locked. The Lagrange-multiplier condition~\cite{Niemi2026} then reduces the 
von Neumann equation to
\begin{equation}
i\hbar\dot\rho_*=[H,\rho_*]=\lambda_a[Q_a,\rho_*],
\label{eq:relative}
\end{equation}
where $\lambda_a$ are the Lagrange multipliers associated with
the prescribed charges $\bm q$ (and 
summation over $a$ is understood). Thus the minimizer evolves 
along the symmetry orbit according to~\cite{Niemi2026} 
\begin{equation}
\rho_*(t)=e^{-i\lambda_aQ_at/\hbar}\rho_*(0)\,e^{i\lambda_aQ_at/\hbar},
\label{time-evo}
\end{equation}
with $\rho_*(0)$ an initial minimizing state. Whenever the r.h.s.\ of Eq.~\eqref{eq:relative}
does not vanish and the evolution~\eqref{time-evo} is periodic, $\rho_*(t)$
realizes a quantum Hamiltonian time crystal~\cite{Niemi2026}; since 
$\rho_*(t)$ is not a Gibbs state the no-go theorem \cite{WatanabeOshikawa2015} does not
apply. 
Since the passive
energy $\ep$ is independent of $\bm q$, varying the prescribed charges at fixed 
spectrum $P$ gives, on differentiable branches of $\emin(P,\bm q)$,
\begin{equation}
d\W_Q(P,\bm q) = \sum_a \lambda_a\,dq_a.
\label{eq:chemical}
\end{equation}

%
%
%
%
%
%

\paragraph*{Spectral selection.}
For a finite-dimensional system with one conserved charge $Q$,
we choose a common eigenbasis of $H$ and $Q$, with eigenvalues
$E_j$ and $m_j$. The Schur--Horn theorem~\cite{Horn1954}
implies that the diagonal entries of $\rho$, viewed as a vector,
range over the convex hull of all permutations
of the fixed spectrum $P$.
Since $q(\rho)$ and $E(\rho)$ depend linearly
on these entries, the pairs $(q(\rho),E(\rho))$ form a
convex polygon as $\rho$ varies over $\Op$.
At fixed spectrum $P$, the lower boundary of this polygon
is the graph of $\emin(P,q)$. Hence $\emin(P,q)$ and $\W_Q(P,q)$
are convex, piecewise-linear functions of the prescribed charge $q$ 
with $\lambda=\partial_q\W_Q$ a constant on each
linear branch.

The constrained-minimum condition~\eqref{eq:relative} implies
\begin{equation}
\bigl[E_j-E_k-\lambda(m_j-m_k)\bigr] {\rho_*}_{jk}=0.
\label{eq:gap}
\end{equation}
Thus a nonzero coherence between different charge sectors
requires $\lambda=(E_j-E_k)/(m_j-m_k)$, corresponding to a
spectral crossing of $H-\lambda Q$.
These crossings determine the possible plateau values
of the symmetry-rotation rate $\lambda/\hbar$,
while the spectrum $P$ determines which plateaus occur
and their intervals in $q$.

%
%
%
%
%
%

\paragraph*{Covariant relaxation.}
With the covariant derivative
$D_t\rho=\dot\rho+(i/\hbar)\lambda_a[Q_a,\rho]$,
the covariant generalization~\cite{Niemi2026} of the qLLG equation~\cite{Liu2024} reads
\begin{equation}
i\hbar D_t\rho=
[H-\lambda_a Q_a+\hbar\kappa D_t\rho,\rho],
\qquad \kappa>0.
\label{eq:qLLG}
\end{equation}
It preserves the density-matrix spectrum and, for time-independent $H$ and $Q_a$, it gives
\begin{equation}
\dot E=\lambda_a\dot q_a-\diss,
\qquad \diss\equiv\kappa\hbar\Tr[(D_t\rho)^2]\geq0.
\label{eq:balance}
\end{equation}
The first term accounts for energy changes associated with charge transfer,
the second for the energy lost through relaxation.
Choosing the multipliers $\lambda_a(t)$ to enforce $\dot q_a=0$
makes the energy decrease monotonically, $\dot E=-\diss\leq0$.
If the flow converges to the global constrained minimum, the total energy loss
equals $\W_{\rm free}(\rho_0)$ of the initial state. 
A state at the global constrained energy minimum may still have
$\dot\rho_*\neq 0$ while $\diss=0$ (\ref{time-evo}), with fixed charge
expectations preventing release of the charge-locked
ergotropy $\W_Q(P,\bm q)$.
Identifying $\diss$ with heat and 
$-\lambda_a\dot q_a$ with useful output power, requires an explicit model for bath and load.

%
%
%
%
%
%

\paragraph*{Two Josephson-coupled islands.}
We consider a quantum battery of two identical superconducting charge islands coupled by a Josephson junction.
In the standard charge--phase description~\cite{Josephson1962,Devoret2004,Rasmussen2021,Koch2007},
with zero offset charges and negligible mutual capacitance, the Hamiltonian is
\begin{equation}
H=\Ec\bigl(N_1^2+N_2^2\bigr)
-\frac{E_J}{2}\bigl(T_1T_2^\dagger+T_1^\dagger T_2\bigr),
\qquad E_J>0,
\label{eq:pair}
\end{equation}
where $\Ec=(2e)^2/(2C_\Sigma)$ is the charging-energy coefficient and $C_\Sigma$
is the capacitance of each island. The operator $N_i$ counts excess Cooper pairs
on island $i$, and the charge-shift operators satisfy
$T_i\ket{m}_i=\ket{m+1}_i$, with $m\in\mathbb Z$.
Thus $T_iT_j^\dagger$ transfers one Cooper pair from island $j$ to island $i$.
Tunneling preserves the sum $Q=N_1+N_2$, $[H,Q]=0$, and
the Hamiltonian is block diagonal in sectors of definite total
charge. Introducing the relative charge $n=(N_1-N_2)/2$
and with $T=T_1T_2^\dagger$,
the Hamiltonian in the total-charge sector $Q=M\in\mathbb Z$ is
\begin{equation}
H_M=\tfrac12\Ec M^2+2\Ec n^2
-\tfrac12 E_J(T+T^\dagger).
\label{eq:sector}
\end{equation}
Within this eigenspace, $n$ has integer eigenvalues for even $M$
and half-integer eigenvalues for odd $M$.
Although $H$ conserves $Q$, a state may retain coherence
between different total-charge sectors. This coherence
underlies the evolution~\eqref{time-evo}~\cite{SM}
and is defined relative to a superconducting phase reference.
Related issues of phase observability arise in superfluids~\cite{Volovik2013},
where coherent dynamics between two time crystals has been
observed~\cite{Autti2021}.

%
%
%
%
%
%

\paragraph*{Exact charge recurrence.}
On the full integer charge lattice, the unitary joint shift operator $B=T_1T_2$
raises the Cooper-pair number of each island by one. It therefore maps
the total-charge eigenspace labeled by $M$ onto that labeled by $M+2$,
while leaving the relative charge $n$ intact. Since $B$ also commutes with
the Josephson term, we have $ B^\dagger H B=H+2\Ec(Q+1) $ and $
B^\dagger Q B=Q+2.$ Thus, for a density matrix $\rho$, the shifted state $\rho'=B\rho B^\dagger$
has the same spectrum, with $q(\rho')=q(\rho)+2$ and
$E(\rho')=E(\rho)+2\Ec[q(\rho)+1]$. This invertible map relates one-to-one
the admissible states at mean charges $q$ and
$q+2$ on the same isospectral orbit, so the constrained minimum energies
satisfy the exact recurrence
\begin{equation}
\emin(P,q+2)=\emin(P,q)+2\Ec(q+1).
\label{eq:recursion}
\end{equation}
Since $E$ and $\emin$ shift equally,
$\W_{\rm free}(B\rho B^\dagger)=\W_{\rm free}(\rho)$, and for every state
and every $E_J$ the joint shift adds only locked energy.

%
%
%
%
%
%

\paragraph*{Even charges.}
We choose $P$ to be the spectrum of the Gibbs state
$\rho_G=e^{-\beta H}/Z$. It 
describes equilibrium with a Cooper-pair reservoir
at the reference chemical potential~\cite{SM}.
The invariance of $H$ under simultaneous charge inversion
$\ket{m_1,m_2}\mapsto\ket{-m_1,-m_2}$ implies
$q(\rho_G)=0$.
The Gibbs state is also passive~\cite{Pusz1978},
so no unitary transformation can lower its energy.
It therefore attains the constrained minimum at $q=0$,
giving $\emin(P,0)=\ep=E_G$, where $E_G=E(\rho_G)$.
Iterating~\eqref{eq:recursion} then gives, for every integer $k$ (with $q=2k$), every 
$\beta>0$, and every $E_J$ the charge-locked ergotropy
\begin{equation}
\W_Q(P,2k)=\emin(P,2k)-E_G
=2\Ec k^2=\tfrac12\Ec q^2.
\label{eq:even}
\end{equation}
This coincides with the electrostatic energy that is added by shifting both
island charges by $k$. The constrained minimum at $q=2k$
is the boosted Gibbs state
$B^k\rho_GB^{-k}\propto \exp\{ -\beta(H-2k\Ec Q) \} $. It describes the equilibrium state at
chemical potential $2k\Ec$. The energy locked in this state remains inaccessible as work
even in contact with a bath at the reference temperature,
provided the pair's total charge is conserved.

\begin{figure}[t]
\centering
\includegraphics[width=0.9\columnwidth]{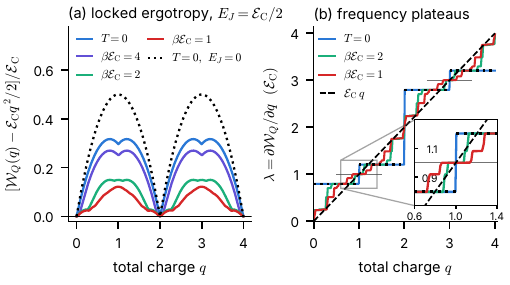}
\caption{Spectral structure of the pair at $E_J=\Ec/2$.
(a) $\W_Q(P,q)-\Ec q^2/2$ for pure and Gibbs spectra:
the two-unit recurrence is exact, while the curves between
even mean charges depend on reference temperature.
The dotted curve and the
zero line are the bounds~\eqref{eq:bounds}.
(b) Slopes $\lambda=\partial\W_Q/\partial q$, showing the
Josephson-split branches $\Ec(2k+1)\mp\delta_J$ (dotted),
the pinned plateaus $\Ec(2k+1)$ (gray),
and further thermal segments; the dashed line is $\Ec q$.
}
\label{fig:pair}
\end{figure}

More generally, the recurrence implies that
$\W_Q(P,q)-\tfrac12\Ec q^2$ is periodic in $q$ with
period $2$ [Fig.~\ref{fig:pair}(a)].
Consequently, $\lambda(q+2)=\lambda(q)+2\Ec$ on
differentiable branches. However, even though~\eqref{eq:even} fixes the 
locked ergotropy  at every even integer value $q=2k$, the recurrence does
not determine the curve $\W_Q(P,q)$ between these values. 
Together with convexity it gives an
upper bound, and a Gibbs variational argument gives
the lower bound~\cite{SM} so that
for $2k\le q\le2k+2$, $\beta>0$, and $E_J\ge0$,
\begin{equation}
0\le\W_Q(P,q)-\tfrac12\Ec q^2
\le\tfrac12\Ec\bigl[1-(q-2k-1)^2\bigr].
\label{eq:bounds}
\end{equation}

%
%
%
%
%
%

\paragraph*{Spectral staircase and parity splitting.}
To determine the locked ergotropy between even mean charges,
we consider the relative Hamiltonian
$H_M-\tfrac12\Ec M^2$ in ~\eqref{eq:sector}.
It depends on the total-charge eigenvalue $M$ only 
through the allowed values of the relative charge
$n$ which is integer for even $M$ and half-integer
for odd $M$.  
The lowest energy at fixed total charge $M$ therefore
has the form
$E_0(M)=\tfrac12\Ec M^2+e_0^{\mathrm{parity}(M)}$,
where $e_0^{\rm even}$ and $e_0^{\rm odd}$ are the lowest
energies of the relative Hamiltonian on these integer
and half-integer ladders, respectively.
For $E_J>0$, the odd-charge ground energy lies below the
straight line joining the neighboring even-charge ground
energies by
\begin{equation}
\begin{aligned}
\delta_J
&=\frac{E_0(2k)+E_0(2k+2)}{2}-E_0(2k+1)\\
&=\tfrac12\Ec-\bigl(e_0^{\rm odd}-e_0^{\rm even}\bigr).
\end{aligned}
\label{eq:deltaJ}
\end{equation}
A variational comparison of the
integer and half-integer relative-charge ladders establishes
$0<\delta_J<\tfrac12\Ec$ for every finite $E_J>0$.
At weak coupling, tunneling mixes the degenerate
$n=\pm\tfrac12$ states, delocalizing the extra Cooper
pair between the islands and giving
$\delta_J\simeq E_J/2-3E_J^2/(16\Ec)$~\cite{SM}.


In the zero-temperature limit, the reference Gibbs state
becomes pure. On its unitary orbit, $\emin(P,q)$ is the
lower convex envelope of the points $(M,E_0(M))$.
For $E_J>0$, the bounds on $\delta_J$ ensure that the successive
differences $E_0(M+1)-E_0(M)$ increase strictly with $M$, so the
envelope joins adjacent points without skipping any, with
alternating slopes~\cite{SM}
\begin{equation}
\lambda=
\begin{cases}
\Ec(2k+1)-\delta_J, & 2k<q<2k+1,\\
\Ec(2k+1)+\delta_J, & 2k+1<q<2k+2.
\end{cases}
\label{eq:slopes}
\end{equation}
For $M<q<M+1$, the minimizing pure state is a coherent
superposition of the lowest-energy states at total charges
$M$ and $M+1$, with probabilities $M+1-q$ and $q-M$~\cite{SM}.
Since their charges differ by one, Eq.~\eqref{eq:gap} gives
$E_0(M+1)-E_0(M)=\lambda$.
Josephson coupling therefore produces parity-split plateaus
in $|\lambda|/\hbar$, the Josephson frequency of the pair
relative to the reference [Fig.~\ref{fig:pair}(b)].
Up to an additive constant, $H-\lambda Q$ is the
Hamiltonian with gate-induced offset charge
$\lambda/(2\Ec)$ in Cooper-pair units, on each island.
Each plateau value of $\lambda$ makes the two states
degenerate in this shifted Hamiltonian, so at zero temperature
$\lambda(q)$ is the inverse of its Coulomb staircase
$q(\lambda)$~\cite{Bouchiat1998}.
At an integer endpoint $q=M$, the superposition reduces
to the lowest-energy eigenstate of charge $M$, which is stationary.

At finite temperature, crossings involving excited
levels of $H-\lambda Q$ satisfying~\eqref{eq:gap}
generate additional linear branches of $\emin(P,q)$ [Fig.~\ref{fig:pair}(a,b)].
Passive ordering assigns the fixed eigenvalues in $P$, from
largest to smallest, to the levels of $H-\lambda Q$
in increasing order, so the staircase is not thermally rounded.
A crossing between charge sectors contributes a segment
of nonzero width in $q$ only if the assigned
probabilities differ~\cite{SM}.
The principal slopes remain $\Ec(2k+1)\mp\delta_J$, and an
isolated principal crossing spans a charge interval of width
$p_1-p_2$, between the two largest probabilities.
Applying the same crossing condition to a level in
sector $M$ and its image under $B^r$, with $r\ne0$,
gives $\lambda=\Ec(M+r)$ for every $E_J$.
On Gibbs orbits, crossings at odd integer $\lambda/\Ec$
produce plateaus at $\Ec(2k+1)$,
with widths $0.16$ and $0.28$ at
$\beta\Ec=2$ and $1$, respectively, for
$E_J=\Ec/2$ [Fig.~\ref{fig:pair}(b)].
The slope at a prescribed $q$ can therefore change
with temperature, although $\W_Q$ at even
charges remains fixed by~\eqref{eq:even}.


%
%
%
%
%
%

\begin{figure*}[t]
\centering
\includegraphics[width=\textwidth]{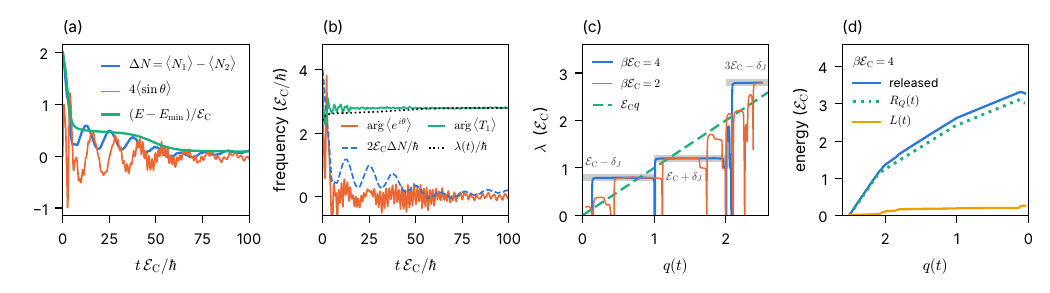}
\caption{Covariant qLLG dynamics, $E_J=\Ec/2$ and $\kappa=1/2$. (a) Charge imbalance $\Delta N$, scaled Josephson
signal $4\langle\sin\theta\rangle$ (both dimensionless) and excess energy $E(t)-\emin(P,q(0))$ in units of $\Ec$, after a
one-sided two-charge boost with $q(0)=2$. (b) Rate of the relative phase for that preparation, with its ac-Josephson
estimate $2\Ec\Delta N/\hbar$ (dashed), compared with the rate of the common phase of a separately prepared $q=2.5$
state and its multiplier $\lambda(t)/\hbar$ (dotted); $\beta\Ec=2$ in (a,b). (c) Multiplier during a prescribed leak with
$q(0)=2.5$, at $\gamma=2.5\times10^{-4}\,\Ec/\hbar$ and the indicated temperatures. The gray bands mark the
parity-split plateaus $\Ec(2k+1)\mp\delta_J$ of Eq.~\eqref{eq:slopes}, labeled by their values; the green dashed line is
the classical $\Ec q$. The negative excursions in the transition windows are clipped.
(d) Released energy $E_0-E(t)$, charge-transfer contribution $R_Q(t)$ (dotted) and internal loss $L(t)$ along the
$\beta\Ec=4$ leak of (c); $|m_i|\le4$ in (c,d).}
\label{fig:dyn}
\end{figure*}

\paragraph*{Locking and the moving minimum.}
With $t=0$ marking the end of preparation, 
Figure~\ref{fig:dyn}(a,b) compares two fixed-charge relaxations
at $E_J=\Ec/2$, $\kappa=1/2$, and  temperature
$\beta\Ec=2$, 
The preparation at $q(0)=2$ provides an exact stationary
energy benchmark, whereas that at $q(0)=2.5$ relaxes toward a
moving minimum with a predicted frequency.

The initial state with $q(0)=2$ is prepared from the Gibbs state by the one-sided boost
$\rho_0=T_1^2\rho_GT_1^{-2}$, giving $E(\rho_0)=E_G+4\Ec$.
Josephson exchange redistributes charge between the islands, while qLLG damping removes energy.
Numerical integration of (\ref{eq:qLLG})
at fixed $q(t)=2$ shows a decaying charge-imbalance beat
as the energy decreases toward that of the stationary
balanced minimum $\rho_*=B\rho_GB^\dagger$, with
$E(\rho_*)=E_G+2\Ec$ [Fig.~\ref{fig:dyn}(a)].
The available fixed-charge relaxation energy is therefore
$\W_{\rm free}(\rho_0)=2\Ec$, exactly half the added charging energy, with the remaining $2\Ec$ charge-locked.
This is a quantum, finite-temperature counterpart of the
two-capacitor problem~\cite{Zucker1955,Mita1999,Boykin2002,SM},
with free ergotropy corresponding to the energy dissipated
in the classical circuit.
Both the one-sided boost $T_1^2$ and the balanced boost
$B=T_1T_2$ add $2\Ec$ to $\tfrac12\Ec\langle Q^2\rangle$,
leaving the Josephson energy unchanged.
The one-sided boost also shifts $n\mapsto n+1$, adding
another $2\Ec$ of relative-charge energy.
This excess can be extracted by $U=T_1^\dagger T_2$, which maps
$\rho_0$ to $\rho_*$.

For the second relaxation, we start from the one-sided
boosted Gibbs state at $q=2$ and prepare a state with mean
charge $q(0)=2.5$ through temporary coupling to an external
superconducting phase reference~\cite{SM}.
After disconnecting the reference at $t=0$, the state
$\rho_0\in\Op$ retains coherence between total-charge
sectors, with constrained minimum on the
branch with slope $\lambda=3\Ec-\delta_J$.
At the exact moving minimum ~\eqref{time-evo}
$[Q,T_i]=T_i$ imply that the phases of
$\langle T_1\rangle$ and $\langle T_2\rangle$ advance
at the same rate $\lambda/\hbar$, while
$\langle T_1T_2^\dagger\rangle$ remains  constant;
in the numerical relaxation it  is
approximately stationary, and the measured common-phase
rate $2.79\,\Ec/\hbar$ agrees within $0.3\%$ with the
spectral prediction
$(3\Ec-\delta_J)/\hbar\simeq2.796\,\Ec/\hbar$
[Fig.~\ref{fig:dyn}(b)]~\cite{SM}.
%
%
%
%
%
%

%
%
%
%
%
%

\paragraph*{Discharge.}

We probe charge release from the prepared $q(0)=2.5$
state by prescribing a slow mean-charge leak
$\dot q(t)=-\gamma<0$.
At each instant, the covariant qLLG charge current
satisfies
$\dot q=\mathcal M(\rho)[\lambda-\lambda_0(\rho)]$,
where $\lambda_0(\rho)$ is the multiplier that would enforce zero
charge current at the same state, and the state-dependent
mobility, derived from (\ref{eq:qLLG}) is~\cite{SM}
\begin{equation}
\mathcal M(\rho)=
\frac{\kappa}{\hbar}
\sum_{j,k}
\frac{(p_j-p_k)^2 |Q_{jk}|^2}
     {1+\kappa^2(p_j-p_k)^2}
\geq 0 .
\label{eq:mobility}
\end{equation}
Here $p_j$ are the eigenvalues of $\rho$, and
$Q_{jk}=\langle j|Q|k\rangle$ is evaluated in an instantaneous
eigenbasis $\rho|j\rangle=p_j|j\rangle$.
The mobility vanishes precisely when $[Q,\rho]=0$
in which case no finite multiplier can produce a charge current.

At a constrained minimum on a differentiable branch,
$\lambda_0$ equals the spectral slope $\partial_q\emin(P,q)$.
When the state tracks a minimum-energy branch and the
offset $\gamma/\mathcal M$ is small,
$\lambda(t)$ stays close to this slope. 
The slow-leak simulations approximately track the
parity-split plateaus away from transition regions
[Fig.~\ref{fig:dyn}(c)].
For $\beta\Ec=4$, the plateau deficits agree within $8\%$
with a branch estimate of the time-averaged offset
$\gamma/\mathcal M$ over the same intervals~\cite{SM}.
The spectral slope minus $\Ec q$ has period $2$ in $q$.
At current $I=2e\gamma$, $\lambda(t)$ approximately follows
this modulation at frequency $I/(4e)$.

Equation~\eqref{eq:balance} gives
\[
E(t)
= E_0 - \gamma \!  \! \int\limits_0^t  \!  \lambda(t')  \dd t' - \int\limits_0^t  
\!  \diss(t') \dd t'  \equiv E_0  - R_Q(t)-L(t),
\]
separating the net released energy into the signed
charge-transfer contribution
and the nonnegative internal loss
[Fig.~\ref{fig:dyn}(d)].
On a minimum-energy branch, the spectral slope acts
as the open-circuit potential of the battery and 
$\lambda$ plays the role of the terminal potential,
lowered by $\gamma/\mathcal M$.
At the constrained minimum, the corresponding
internal dissipation rate is
$\diss=\gamma^2/\mathcal M$.
The two simulated discharges give internal-loss fractions
of approximately $8\%$ and $16\%$ at $\beta\Ec=4$ and $2$,
respectively~\cite{SM}.
Figure~\ref{fig:dyn}(d) shows the colder run.
Since $\mathcal M(\rho)=0$ when $[Q,\rho]=0$,
no finite multiplier can sustain the prescribed
current at a charge-diagonal state.

%
%
%
%
%
%

\vskip 0.1cm

\paragraph*{Summary.}
We have investigated how fixing the mean total charge can
lock part of the ergotropy in a quantum battery of two
Josephson-coupled superconducting islands.
Starting from a thermal state, adding one Cooper pair to
each island locks all the added energy.
Adding two Cooper pairs to one island costs twice as much
energy but locks the same amount, leaving exactly half
extractable by controlled charge-preserving unitaries
and reproducing the classical two-capacitor energy partition.
Both results hold at every temperature and Josephson coupling.
The balanced even-charge minima are equilibrium states at a
shifted chemical potential, and therefore stationary.
At other mean charges, minimizing states can sustain
coherent phase evolution relative to a phase reference,
without external driving or consumption of stored energy,
at frequencies that Josephson coupling splits by charge parity.
A prescribed-current qLLG discharge model separates
charge-transfer energy from internal dissipation.
The supplement outlines a possible circuit
realization~\cite{SM}.
Determining how much is captured as useful work requires
an explicit receiver model, beyond this phenomenological
description.

%
%
%
%
%
%
%

\vspace{0.3cm}
\paragraph*{Acknowledgments}
This research is supported by the Swedish Research Council under 
Contract No.~2022-04037 and by COST Action CA23134 (POLYTOPO).

The simulation codes were written with the 
assistance of AI tools (both Anthropic Claude and OpenAI models) 
under the author's direction. All code, analyses
and figures were checked by the author, who takes full responsibility for them, and
all data is available at \url{https://github.com/niemiantti/Battery}.
AI tools were also used in support of
theoretical analyses and for language  editing of the manuscript.


\begin{thebibliography}{28}
\urlstyle{same}
\bibitem{Alicki2013} R. Alicki and M. Fannes, Entanglement boost for extractable work from ensembles of
quantum batteries, Phys. Rev. E \textbf{87}, 042123 (2013),
\url{https://doi.org/10.1103/PhysRevE.87.042123}.

\bibitem{Ferraro2018} D. Ferraro, M. Campisi, G. M. Andolina, V. Pellegrini, and M. Polini,
High-power collective charging of a solid-state quantum battery,
Phys. Rev. Lett. \textbf{120}, 117702 (2018),
\url{https://doi.org/10.1103/PhysRevLett.120.117702}.

\bibitem{Paulino2026} P. J. Paulino, A. Cabot, G. De Chiara, M. Antezza, I. Lesanovsky, and F. Carollo,
Thermodynamics of coupled time crystals with an application to energy storage,
Quantum Sci. Technol. \textbf{11}, 015003 (2026),
\url{https://doi.org/10.1088/2058-9565/ae186c}.

\bibitem{Binder2015} F. C. Binder, S. Vinjanampathy, K. Modi, and J. Goold, Quantacell: Powerful charging
of quantum batteries, New J. Phys. \textbf{17}, 075015 (2015),
\url{https://doi.org/10.1088/1367-2630/17/7/075015}.

\bibitem{Campaioli2017} F. Campaioli, F. A. Pollock, F. C. Binder, L. C\'eleri, J. Goold,
S. Vinjanampathy, and K. Modi, Enhancing the charging power of quantum batteries,
Phys. Rev. Lett. \textbf{118}, 150601 (2017),
\url{https://doi.org/10.1103/PhysRevLett.118.150601}.

\bibitem{Francica2020} G. Francica, F. C. Binder, G. Guarnieri, M. T. Mitchison, J. Goold,
and F. Plastina, Quantum coherence and ergotropy,
Phys. Rev. Lett. \textbf{125}, 180603 (2020),
\url{https://doi.org/10.1103/PhysRevLett.125.180603}.

\bibitem{Allahverdyan2004} A. E. Allahverdyan, R. Balian, and Th. M. Nieuwenhuizen, Maximal work
extraction from finite quantum systems, Europhys. Lett. \textbf{67}, 565 (2004),
\url{https://doi.org/10.1209/epl/i2004-10101-2}.

\bibitem{Pusz1978} W. Pusz and S. L. Woronowicz, Passive states and KMS states for general quantum
systems, Commun. Math. Phys. \textbf{58}, 273 (1978),
\url{https://doi.org/10.1007/BF01614224};
A. Lenard, Thermodynamical proof of the Gibbs formula for elementary quantum systems,
J. Stat. Phys. \textbf{19}, 575 (1978),
\url{https://doi.org/10.1007/BF01011769}.

\bibitem{Mitsuhashi2022} Y. Mitsuhashi, K. Kaneko, and T. Sagawa, Characterizing symmetry-protected
thermal equilibrium by work extraction, Phys. Rev. X \textbf{12}, 021013 (2022),
\url{https://doi.org/10.1103/PhysRevX.12.021013}.

\bibitem{Hu2022} C.-K. Hu, J. Qiu, P. J. P. Souza, J. Yuan, Y. Zhou, L. Zhang, J. Chu, X. Pan,
L. Hu, J. Li, Y. Xu, Y. Zhong, S. Liu, F. Yan, D. Tan, R. Bachelard, C. J. Villas-Boas,
A. C. Santos, and D. Yu, Optimal charging of a superconducting quantum battery,
Quantum Sci. Technol. \textbf{7}, 045018 (2022),
\url{https://doi.org/10.1088/2058-9565/ac8444}.

\bibitem{Bouchiat1998} V. Bouchiat, D. Vion, P. Joyez, D. Esteve, and M. H. Devoret, Quantum
coherence with a single Cooper pair, Phys. Scr. \textbf{T76}, 165 (1998),
\url{https://doi.org/10.1238/Physica.Topical.076a00165}.

\bibitem{Nakamura1999} Y. Nakamura, Yu. A. Pashkin, and J. S. Tsai, Coherent control of macroscopic
quantum states in a single-Cooper-pair box, Nature \textbf{398}, 786 (1999),
\url{https://doi.org/10.1038/19718}.

\bibitem{Niemi2026} A. J. Niemi, Quantum Hamiltonian time crystals from covariant quantum
Landau--Lifshitz--Gilbert relaxation, arXiv:2610.39886 (2026),
\url{https://doi.org/10.48550/arXiv.2610.39886}.

\bibitem{Zucker1955} C. Zucker, Condenser problem, Am. J. Phys. \textbf{23}, 469 (1955),
\url{https://doi.org/10.1119/1.1934050}.

\bibitem{Mita1999} K. Mita and M. Boufaida, Ideal capacitor circuits and energy conservation,
Am. J. Phys. \textbf{67}, 737 (1999),
\url{https://doi.org/10.1119/1.19363}.

\bibitem{Boykin2002} T. B. Boykin, D. Hite, and N. Singh, The two-capacitor problem with radiation,
Am. J. Phys. \textbf{70}, 415 (2002),
\url{https://doi.org/10.1119/1.1435344}.

\bibitem{Hoekstra2020}
J. Hoekstra,
A Solution of the Two-Capacitor Problem Through Its
Similarity to Single-Electron Electronics,
\href{https://doi.org/10.1109/OJCAS.2020.2977216}
{IEEE Open J. Circuits Syst. \textbf{1}, 13--21 (2020)}.


\bibitem{Liu2024} Y. Liu, I. P. Miranda, L. Johnson, A. Bergman, A. Delin, D. Thonig, M. Pereiro,
O. Eriksson, V. Azimi-Mousolou, and E. Sj\"oqvist, Quantum analog of Landau--Lifshitz--Gilbert
dynamics, Phys. Rev. Lett. \textbf{133}, 266704 (2024),
\url{https://doi.org/10.1103/PhysRevLett.133.266704}.

\bibitem{WatanabeOshikawa2015} H.~Watanabe and M.~Oshikawa, Absence of quantum time crystals, Phys. Rev. 
Lett. \textbf{114}, 251603 (2015), \url{https://doi.org/10.1103/PhysRevLett.114.251603}. 


%
%

\bibitem{Horn1954} A. Horn, Doubly stochastic matrices and the diagonal of a rotation matrix,
Am. J. Math. \textbf{76}, 620 (1954),
\url{https://doi.org/10.2307/2372705}.

\bibitem{Josephson1962} B. D. Josephson, Possible new effects in superconductive tunnelling,
Phys. Lett. \textbf{1}, 251 (1962),
\url{https://doi.org/10.1016/0031-9163(62)91369-0}.

\bibitem{Devoret2004} M. H. Devoret, A. Wallraff, and J. M. Martinis, Superconducting qubits: A short review,
arXiv:cond-mat/0411174 (2004),
\url{https://doi.org/10.48550/arXiv.cond-mat/0411174}.

\bibitem{Rasmussen2021} S. E. Rasmussen, K. S. Christensen, S. P. Pedersen, L. B. Kristensen, T. B\ae kkegaard,
N. J. S. Loft, and N. T. Zinner, Superconducting circuit companion---an introduction with worked examples,
PRX Quantum \textbf{2}, 040204 (2021),
\url{https://doi.org/10.1103/PRXQuantum.2.040204}.

\bibitem{Koch2007} J. Koch, T. M. Yu, J. Gambetta, A. A. Houck, D. I. Schuster, J. Majer,
A. Blais, M. H. Devoret, S. M. Girvin, and R. J. Schoelkopf, Charge-insensitive qubit design derived
from the Cooper pair box, Phys. Rev. A \textbf{76}, 042319 (2007),
\url{https://doi.org/10.1103/PhysRevA.76.042319}.

\bibitem{SM} See Supplemental Material for derivations of the spectral results
and details of the state preparation and numerical relaxation and discharge protocols.

\bibitem{Volovik2013} G. E. Volovik, On the broken time translation symmetry in macroscopic 
systems: precessing states and off-diagonal
long-range order, JETP Lett. \textbf{98}, 491 (2013),
\url{https://doi.org/10.1134/S0021364013210133}.

\bibitem{Autti2021} S. Autti, P. J. Heikkinen, J. T. M\"akinen, G. E. Volovik, V. V. Zavjalov, and V. B. Eltsov, AC Josephson effect
between two superfluid time crystals, Nat. Mater. \textbf{20}, 171 (2021),
\url{https://doi.org/10.1038/s41563-020-0780-y}.

\end{thebibliography}
\end{document}


\title{Supplemental Material: \\ \Large Charge-locked ergotropy in a
Josephson-coupled quantum battery}
\author{Antti J. Niemi}
\noaffiliation
\maketitle

\noindent

This supplement provides the derivations and numerical details
supporting the article, and outlines a possible superconducting-circuit
realization for testing its principal predictions.
Each section identifies the corresponding discussion in the article.
Equation and figure numbers without the prefix S refer to the article.

We use the density-matrix spectrum $P$ and unitary orbit $\Op$,
with passive energy $\ep$, constrained minimum $\emin(P,q)$,
and ergotropies $\W_{\rm free}$ and $\W_Q$ defined in
Eqs.~\Ldef\ and~\Lsplit.
The minimizing state $\rho_*$ and multiplier $\lambda$ appear
in Eqs.~\Lstat\ and~\Lorbit; the covariant derivative
$D_t\rho$, damping parameter $\kappa$, and dissipation $\diss$
in Eqs.~\Lqllg\ and~\Lbal.

For the island pair, $Q=N_1+N_2$ counts excess Cooper pairs
and $q=\Tr(\rho Q)$ is its expectation value.
The Hamiltonian $H$, charge-shift operators $T_i$, joint shift
$B=T_1T_2$, and reference Gibbs state $\rho_G$ with energy $E_G$
are introduced around Eqs.~\Lham--\Leven.

\section{Reference state and phase reference}
\label{sec:ref}
\supports{``Two Josephson-coupled islands'' and ``Even charges'': the phase
variables, the reference state, and the phase reference for coherence
between total-charge sectors.}

\textit{Phase variables.}
The charge-shift operator raises the excess Cooper-pair number,
$T_i\ket{m_i}=\ket{m_i+1}$, with $m_i\in\mathbb Z$.
In the phase representation, $\psi(\varphi_1,\varphi_2)$ is
$2\pi$-periodic in each coordinate, and the operators act as
$N_i\psi=-i\partial_{\varphi_i}\psi$ and $T_i\psi=e^{i\varphi_i}\psi$.
With the relative phase $\theta=\varphi_1-\varphi_2$, the
relative charge $n=(N_1-N_2)/2$ acts as $-i\partial_\theta$ and
$T=T_1T_2^\dagger$ as multiplication by $e^{i\theta}$.
At total charge $Q=M$, Eq.~\Lsector\ gives the relative Hamiltonian
$H_M-\tfrac12\Ec M^2=2\Ec n^2-E_J\cos\theta$.
Its wavefunctions satisfy
$f_M(\theta+2\pi)=(-1)^M f_M(\theta)$, giving integer eigenvalues
of $n$ for even $M$ and half-integer values for odd $M$.

\textit{Reference state.}
Let $\Pi_M$ project onto the sector $Q=M$, and let
$w_M(\rho)=\Tr(\Pi_M\rho)$ denote the sector weights of a state $\rho$.
The reference Gibbs state at inverse temperature
$\beta>0$ is a mixture of normalized thermal states
$\sigma_M$ at fixed total charge $M$:
\begin{equation}
\rho_G=\frac{e^{-\beta H}}{Z}
=\bigoplus_M w_M^G\,\sigma_M,\qquad
\sigma_M=\frac{e^{-\beta H_M}}{Z_M}, \qquad
w_M^G\equiv w_M(\rho_G)=\frac{Z_M}{Z},\qquad
Z_M=e^{-\beta\Ec M^2/2}\,z_{\mathrm{parity}(M)}.
\label{eq:blocks}
\end{equation}
Here $Z=\sum_M Z_M$ and
$z_{\rm even/odd}=\sum_\ell e^{-\beta e_\ell^{\rm even/odd}}$,
where $e_\ell^{\rm even/odd}$, with $\ell=0,1,\ldots$, are the eigenvalues
of the relative Hamiltonian for even and odd $M$,
respectively.
Equation~\eqref{eq:blocks} describes equilibrium
with a Cooper-pair reservoir at zero reference chemical potential.
Charge exchange with the reservoir gives the sector weights $w_M^G$.
Charge inversion implies $w_{-M}^G=w_M^G$ and hence
$q(\rho_G)=\Tr(Q\rho_G)=0$.
If the total charge is fixed at $M$, the thermal state is instead
$\sigma_M$, with a different density-matrix spectrum.
The recurrence~\Lrec\ holds for either spectrum.
The equilibrium interpretation of the even-charge minima
and the lower bound~\Lbounds\ use the full Gibbs spectrum.

\textit{Phase reference.}
A nonzero $\langle T_i\rangle$ requires coherence
between total-charge sectors (intersector coherence), since it vanishes
when $[Q,\rho]=0$.
Its phase is defined relative to an external superconducting
phase reference~\cite{Bartlett2007}, labeled $L$, with charge-shift
operator $T_L$.
The operator $T_iT_L^\dagger$ transfers a Cooper pair
from the reference to island $i$ at fixed combined charge; its
expectation value measures their relative phase coherence.
In the fixed-phase approximation for a macroscopic reference, $T_L$ is
replaced by $e^{i\varphi_L}$.
We choose $\varphi_L=0$, so that $\rho$ describes the islands relative to
this phase origin.
The same superconductor can prepare the Gibbs
state~\eqref{eq:blocks} through a weak Josephson contact
(Sec.~\ref{sec:circuit}).
After the contact is removed, $Q$ is conserved and
the reference continues to define the phase.

Equation~\Lorbit\ and $[Q,T_i]=T_i$ give
\begin{equation}
\langle T_i\rangle(t)
=e^{i\lambda t/\hbar}\,\langle T_i\rangle(0).
\label{eq:clock}
\end{equation}
The multiplier $\lambda$ is the islands' Cooper-pair chemical potential
relative to the reference, and Eq.~\eqref{eq:clock} is the corresponding
ac Josephson relation.
When $\langle T_i\rangle\ne0$, a weak Josephson contact
to island $i$ would probe this phase evolution at angular frequency
$|\lambda|/\hbar$.
Since $[Q,T_1T_2^\dagger]=0$, the relative coherence
$\langle T_1T_2^\dagger\rangle$ remains constant
on the orbit~\Lorbit.
A moving minimum satisfies
$[H,\rho_*]=\lambda[Q,\rho_*]\ne0$.
It is neither a ground state nor a Gibbs state of $H$, and thus lies
outside the equilibrium setting of the no-go
theorem~\cite{Watanabe2015}.
Its energy is minimized at fixed spectrum and mean charge.

\section{Even charges and the bounds~\texorpdfstring{\Lbounds}{(13)}}
\label{sec:bounds}
\supports{``Exact charge recurrence'' and ``Even charges'':
the even-charge values~\Leven\ and the bounds~\Lbounds\ [Fig.~1(a)].}

\textit{Even charges.}
The even-charge result~\Leven\ extends to any spectrum $P$ admitting a
finite-energy passive representative.
Simultaneous charge inversion $\ket{m_1,m_2}\mapsto\ket{-m_1,-m_2}$
commutes with $H$ and reverses $Q$.
Each energy eigenspace therefore admits an inversion eigenbasis with zero
mean charge.
Assigning $P$ passively in this basis gives
$\emin(P,0)=\ep$.
The recurrence~\Lrec\ then yields $\W_Q(P,2k)=2\Ec k^2$
for every integer $k$.
Thus the pure and Gibbs spectra of Fig.~1(a) share the even-charge values
and the convexity upper bound~\Lbounds.
The lower bound uses the Gibbs spectrum.

\textit{Upper bound.}
Convexity places $\emin(P,q)$ on or below the chord joining
$\ep+2\Ec k^2$ and $\ep+2\Ec(k+1)^2$ at
$q=2k$ and $q=2k+2$.
Subtracting $\ep+\tfrac12\Ec q^2$ gives the upper bound~\Lbounds.
The bound is saturated on the pure-state orbit at
$E_J=0$, where superpositions of
$\ket{k,k}$ and $\ket{k+1,k+1}$ attain the minimum
[dotted curve in Fig.~1(a)].

\textit{Lower bound.}
Let $P$ be the spectrum of $\rho_G$.
We introduce a common gate offset $x\in\mathbb R$ and define
\begin{equation}
H_x=H-2\Ec xQ+2\Ec x^2
=\Ec\bigl[(N_1-x)^2+(N_2-x)^2\bigr]-E_J\cos\theta,
\qquad Z(x)=\Tr e^{-\beta H_x},
\label{eq:Hx}
\end{equation}
so that $Z(0)=Z$.
Every $\rho\in\Op$ at mean charge $q$ has
$E(\rho)=\Tr(H_x\rho)+2\Ec x(q-x)$ and the von Neumann entropy
$S(\rho)=-\Tr(\rho\ln\rho)=\beta E_G+\ln Z(0)$ of the Gibbs state.
The Gibbs variational principle for $H_x$,
$\Tr(H_x\rho)-\beta^{-1}S(\rho)\ge-\beta^{-1}\ln Z(x)$, therefore gives
\begin{equation}
E(\rho)\ge E_G+2\Ec x(q-x)+\beta^{-1}\ln\frac{Z(0)}{Z(x)} .
\label{eq:lb}
\end{equation}
Using Eq.~\eqref{eq:blocks} and Poisson resummation over $M$,
\begin{equation}
Z(x)=\sum_{m\in\mathbb Z}d_m\bigl[z_{\rm even}+(-1)^mz_{\rm odd}\bigr]\cos(2\pi mx),
\qquad
d_m=\sqrt{\frac{\pi}{2\beta\Ec}}\;e^{-\pi^2m^2/(2\beta\Ec)}>0 .
\label{eq:poisson}
\end{equation}
The even and odd ladders correspond to periodic and antiperiodic
boundary conditions for the operator
$-2\Ec\partial_\theta^2-E_J\cos\theta$ (Sec.~\ref{sec:ref}), whose
heat kernel $K_\beta$ on the line is positive, so that
\[
z_{\rm even}-z_{\rm odd}
=2\sum_{j\,{\rm odd}}\int_0^{2\pi}K_\beta(\theta+2\pi j,\theta)\,\dd\theta>0.
\]
All Fourier coefficients in Eq.~\eqref{eq:poisson} are
therefore nonnegative, and $Z(x)\le Z(0)$.
Setting $x=q/2$ in Eq.~\eqref{eq:lb} and minimizing
over $\rho$ gives
\begin{equation}
\W_Q(P,q)\;\ge\;\tfrac12\Ec q^2+\beta^{-1}\ln\frac{Z(0)}{Z(q/2)}\;\ge\;\tfrac12\Ec q^2 .
\label{eq:lbfinal}
\end{equation}
This is the lower bound in Eq.~\Lbounds.

\section{Parity splitting and the pure-state minima}
\label{sec:parity}
\supports{the bounds on $\delta_J$, its weak-coupling expansion, and the
pure-state slopes~\Lslopes\ in ``Spectral staircase and parity splitting''.}

We write the relative Hamiltonian of Eq.~\Lsector\ as
$h_\alpha=2\Ec n^2-\tfrac12 E_J(T+T^\dagger)$, acting on the
relative-charge states $\ket n$ with $n\in\mathbb Z+\alpha$,
where $\alpha=0$ for even $M$ and $\alpha=\tfrac12$ for odd $M$.
Its lowest eigenvalues are the ground energies $e_0^{\rm even}$ and
$e_0^{\rm odd}$ of the article.

\textit{Strict bounds.}
The ground state $\ket{g_e}$ of $h_0$ is nondegenerate
and even under $n\mapsto-n$, so that
$\langle g_e|n|g_e\rangle=0$.
The half-step shift $\mathsf S\ket n=\ket{n+\tfrac12}$
maps the integer ladder onto the
half-integer ladder and satisfies
\begin{equation}
\mathsf S^\dagger h_{1/2}\mathsf S
=h_0+2\Ec n+\tfrac12\Ec.
\label{eq:halfstep}
\end{equation}
Using $\mathsf S\ket{g_e}$ as a trial state for $h_{1/2}$ therefore gives
$e_0^{\rm odd}\le e_0^{\rm even}+\tfrac12\Ec$.
Equality would require $\mathsf S\ket{g_e}$ to be an eigenstate of
$h_{1/2}$ and hence, by Eq.~\eqref{eq:halfstep}, $n\ket{g_e}=0$.
This is impossible for $E_J>0$, since Josephson
tunneling couples different charge states, and therefore $\delta_J>0$.

For the upper bound we use the
phase representation, where the even and odd
sectors correspond to periodic and antiperiodic
boundary conditions (Sec.~\ref{sec:ref}).
The modulus $|f|$ of the antiperiodic ground-state
wavefunction $f$ is periodic and has the same
potential energy but no greater kinetic energy, so that
$e_0^{\rm even}\le e_0^{\rm odd}$.
Equality would require $|f|$ to coincide with
the unique, strictly positive periodic ground state and $f$ to have a
constant phase, which is incompatible with the
antiperiodic boundary condition.
Consequently,
\begin{equation}
0<\delta_J<\tfrac12\Ec,
\qquad E_J>0.
\label{eq:deltabounds}
\end{equation}

\textit{Weak coupling.}
For $E_J\ll\Ec$, the Josephson term, with matrix element
$-E_J/2$ between neighboring relative-charge states, is a
perturbation of $2\Ec n^2$.
On the integer ladder the unperturbed ground state $\ket{0}$ couples to
$\ket{\pm1}$, which lie $2\Ec$ above it, so that
$e_0^{\rm even}=-E_J^2/(4\Ec)+O(E_J^4/\Ec^3)$.
On the half-integer ladder the degenerate states $\ket{\pm\tfrac12}$, of
energy $\tfrac12\Ec$, are split in first order.
Their bonding combination
$(\ket{\tfrac12}+\ket{-\tfrac12})/\sqrt2$ is lowered by $E_J/2$ and
couples to each of $\ket{\pm\tfrac32}$, which lie $4\Ec$ above the
doublet, with matrix element $-E_J/(2\sqrt2)$, so that
$e_0^{\rm odd}=\tfrac12\Ec-\tfrac12E_J-E_J^2/(16\Ec)+O(E_J^3/\Ec^2)$.
Equation~\LdeltaJ\ then yields
\begin{equation}
\delta_J
=\frac{E_J}{2}-\frac{3E_J^2}{16\Ec}
+O\!\left(\frac{E_J^3}{\Ec^2}\right).
\label{eq:deltaweak}
\end{equation}

\textit{Pure-state minima.}
Let $\ket{g_M}$ denote the ground state of $H_M$, with the
energy $E_0(M)$ of the article.
A pure state $\ket\psi$ with sector weights $w_M=w_M(\proj\psi)$ has
$q=\sum_MMw_M$ and
\begin{equation}
\bra\psi H\ket\psi\ge\sum_Mw_ME_0(M),
\label{eq:sectorbound}
\end{equation}
with equality for $\sum_M\sqrt{w_M}\,e^{i\zeta_M}\ket{g_M}$ and arbitrary
phases $\zeta_M$.
Thus $\emin(P,q)$ is the lower convex envelope of the points $(M,E_0(M))$.
The second differences of $E_0(M)$ are $2(\Ec-\delta_J)$ for even $M$ and
$2\delta_J$ for odd $M$.
Both are positive by Eq.~\eqref{eq:deltabounds}, so that the successive
slopes
\begin{equation}
s_M\equiv E_0(M+1)-E_0(M),\qquad
s_{2k}=\Ec(2k+1)-\delta_J,\qquad
s_{2k+1}=\Ec(2k+1)+\delta_J
\label{eq:slopes}
\end{equation}
increase strictly with $M$.
The lower envelope therefore joins adjacent points, and its
minimizers for $M\le q\le M+1$ are
\begin{equation}
\rho_*=\proj{\psi},\qquad
\ket\psi=\sqrt{M+1-q}\,\ket{g_M}+e^{i\zeta}\sqrt{q-M}\,\ket{g_{M+1}},
\label{eq:purestate}
\end{equation}
with arbitrary phase $\zeta$.
Both components are eigenstates of $H-s_MQ$ with the same eigenvalue,
so that $\rho_*$ satisfies the stationarity condition~\Lstat\ with
$\lambda=s_M$.
This gives the slopes~\Lslopes\ and the probabilities quoted
in the article.
Under Eq.~\Lorbit, the phase $\zeta$ evolves as
$\zeta(t)=\zeta(0)-s_Mt/\hbar$.

\section{Finite-temperature staircase}
\label{sec:thermal}
\supports{passive ordering, plateau widths, pinned plateaus,
and the Coulomb-staircase comparison in
``Spectral staircase and parity splitting'' [Fig.~1], and the
modulation frequency $I/4e$ in ``Discharge''.}

\textit{Passive bound.}
Let $E_j$ and $m_j$ denote the eigenvalues
of $H$ and $Q$ in a common eigenbasis, as in the article.
For each $\lambda$, we arrange the shifted energies
$E_j-\lambda m_j$ in increasing order,
$\eta_{(1)}(\lambda)\le\eta_{(2)}(\lambda)\le\cdots$,
and the fixed spectrum $P$ in decreasing order,
$p_1\ge p_2\ge\cdots$.
Assigning the largest probabilities to the lowest shifted energies
minimizes the expectation value of $H-\lambda Q$ on $\Op$:
\begin{equation}
\Phi_P(\lambda)
\equiv\min_{\rho\in\Op}\Tr[(H-\lambda Q)\rho]
=\sum_r p_r\,\eta_{(r)}(\lambda).
\label{eq:passive}
\end{equation}
Every state $\rho\in\Op$ with mean charge $q$ therefore
satisfies $E(\rho)\ge\Phi_P(\lambda)+\lambda q$
for all $\lambda$, with equality only if $\rho$ is passive with respect to
$H-\lambda Q$.
Such a state commutes with $H-\lambda Q$, which is the
stationarity condition~\Lstat\ with multiplier $\lambda$,
and its coherences obey Eq.~\Lgap.
For the pure and Gibbs spectra used here, equality is attained at every
$q$ by a passive assignment or a rotation within a degenerate eigenspace,
so that
\begin{equation}
\emin(P,q)
=\sup_\lambda\bigl[\Phi_P(\lambda)+\lambda q\bigr].
\label{eq:dual}
\end{equation}
Quadratic charge confinement and the decay of the Gibbs probabilities
ensure convergence of the sum~\eqref{eq:passive} on the full lattice.
Equation~\eqref{eq:dual} extends the article's finite-dimensional
convexity argument to these spectra.

\textit{Crossings and widths.}
The passive ordering changes when two shifted energies
cross as $\lambda$ varies.
Consider an isolated crossing between levels $j$ and $k$,
with $m_k>m_j$, at $\lambda_*=(E_k-E_j)/(m_k-m_j)$.
Immediately below the crossing, level $j$ lies below
level $k$ in the shifted-energy ordering, and the two
levels carry the probabilities $p_r\ge p_{r+1}$ of
consecutive ranks; let $q_-$ and $E_-$ denote the charge and the energy of
this passive assignment.
At $\lambda=\lambda_*$ the two levels are degenerate
with respect to $H-\lambda_*Q$, and the density matrix
can be rotated within their two-dimensional subspace
without changing its shifted energy.
Keeping all other probabilities unchanged, the rotated state assigns
$p_r$ and $p_{r+1}$ to
\begin{equation}
\ket{\chi_1}
=\cos\vartheta\,\ket{j}
 +e^{i\zeta}\sin\vartheta\,\ket{k},
\qquad
\ket{\chi_2}
=-\sin\vartheta\,\ket{j}
 +e^{i\zeta}\cos\vartheta\,\ket{k},
\qquad 0\le\vartheta\le\tfrac{\pi}{2}.
\label{eq:ab}
\end{equation}
The rotation preserves $P$ and $\Phi_P(\lambda_*)$, so the
rotated state attains
the constrained minimum at its own mean charge.
For $0<\vartheta<\pi/2$ it carries the coherence
$\bra{j}\rho\ket{k}=(p_r-p_{r+1})e^{-i\zeta}\sin\vartheta\cos\vartheta$
between the charge sectors $m_j$ and $m_k$,
whose phase advances according to Eq.~\Lorbit.
Its mean charge is
$q(\vartheta)=q_-+(p_r-p_{r+1})(m_k-m_j)\sin^2\vartheta$, so that the
rotation from $\vartheta=0$ to $\pi/2$ covers
\begin{equation}
q_-\le q\le q_+=q_-+(p_r-p_{r+1})(m_k-m_j),
\qquad
\emin(P,q)=E_-+\lambda_*(q-q_-).
\label{eq:width}
\end{equation}
The crossing therefore generates a linear branch
with slope $\lambda_*$, of nonzero width only if the
two assigned probabilities differ, as stated in the article.
These probabilities are assigned by rank in
the passive ordering, not by the identities
of the crossing levels.
For simultaneous crossings in disjoint pairs of levels, the widths add.

\textit{Coulomb staircase.}
Away from crossings, the passive assignment has
mean charge
$q_P(\lambda)=-\partial_\lambda\Phi_P(\lambda)
=\sum_r p_r m_{(r)}(\lambda)$,
where $m_{(r)}(\lambda)$ is the charge of the level
occupying rank $r$ in the shifted-energy ordering.
This charge is constant between crossings and jumps from
$q_-$ to $q_+$ at $\lambda_*$, so that the slope
$\lambda(q)=\partial_q\emin(P,q)$ is the inverse of the staircase
$q_P(\lambda)$.
For a pure spectrum
$\Phi_P(\lambda)=\min_M[E_0(M)-\lambda M]$,
and $q_P(\lambda)$ is the zero-temperature
Coulomb staircase of the pair (cf.~Ref.~\cite{Bouchiat1998}) at the gate offset
$x=\lambda/(2\Ec)$ of Eq.~\eqref{eq:Hx}.
For the Gibbs spectrum the eigenvalues $p_r$ remain
fixed as $\lambda$ varies and only their assignment to
the levels changes, so that $q_P(\lambda)$ remains a staircase.
Thermal equilibrium with $H-\lambda Q$ at each $\lambda$ would instead
assign Boltzmann weights to the shifted energies and give the thermally
rounded charge
$q_{\rm eq}(\lambda)=\Tr[Qe^{-\beta(H-\lambda Q)}]/\Tr e^{-\beta(H-\lambda Q)}$;
the two coincide at $\lambda=2k\Ec$, where the passive state is
$B^k\rho_GB^{-k}\propto e^{-\beta(H-2k\Ec Q)}$.

\textit{Principal plateaus.}
The shifted energies $E_0(M)-\lambda M$ of neighboring
sector ground states cross at $\lambda=s_M$
[Eq.~\eqref{eq:slopes}].
Since $s_M$ increases strictly with $M$, at
$\lambda=s_M$ the states $\ket{g_M}$ and $\ket{g_{M+1}}$
form the degenerate ground level of $H-\lambda Q$ and occupy the
ranks $1$ and $2$.
Their charges differ by one.
For an isolated principal crossing, Eq.~\eqref{eq:width} gives the width
and endpoints,
\begin{equation}
\Delta q_{\rm pr}=p_1-p_2,
\qquad
q_-=p_1M+p_2(M+1)
  +\sum_{r\ge3}p_r m_{(r)}(s_M^-),
\qquad
q_+=q_-+p_1-p_2.
\label{eq:principal}
\end{equation}
Here $m_{(r)}(s_M^-)$ is evaluated immediately below the
crossing.
The two largest probabilities determine
the width, while the full passive assignment
determines the position of the interval.
For the Gibbs spectrum, the first excited level of $H$ consists of
$\ket{g_1}$ and $\ket{g_{-1}}$, at the excitation energy
$s_0=\Ec-\delta_J$.
Hence $p_2=p_3=p_1e^{-\beta(\Ec-\delta_J)}$ and
$\Delta q_{\rm pr}=p_1[1-e^{-\beta(\Ec-\delta_J)}]$,
where $p_1$ is the ground-state probability in $\rho_G$.
As $\beta\to\infty$ the principal interval approaches $[M,M+1]$, the
pure-state branch of Sec.~\ref{sec:parity}.

\textit{Pinned plateaus.}
The shift $B^r$, with nonzero integer $r$, maps a level
in sector $M$ to the level with the same relative state
in sector $M+2r$.
Their charges differ by $2r$ and, by the relation
$B^\dagger HB=H+2\Ec(Q+1)$ of the article,
their energies by $2r\Ec(M+r)$.
The two shifted energies therefore cross at
\begin{equation}
\lambda=\Ec(M+r)
\label{eq:pin}
\end{equation}
for every $E_J$.
A crossing opens a branch only if its two ranks
carry unequal probabilities.
At $\lambda=2k\Ec$ the passive state is
$B^k\rho_GB^{-k}$, whose probabilities are equal within
every degenerate level of $H-\lambda Q$, and no branch opens.
At $\lambda=(2k+1)\Ec$ the levels of the mirrored sectors
$M$ and $4k+2-M$ are degenerate in pairs whose
assigned probabilities generally differ; each such pair contributes a
width given by Eq.~\eqref{eq:width}, and the contributions of disjoint
pairs add.
The resulting plateau is pinned to the odd integer
value of $\lambda/\Ec$, while its width depends
on $E_J$ and on the Gibbs spectrum.
For $E_J=\Ec/2$, the lowest mirrored pair,
$\ket{g_{2k}}$ and $\ket{g_{2k+2}}$, occupies ranks $2$ and $3$
and contributes no width because $p_2=p_3$.
Higher pairs give the total widths in Table~\ref{tab:thermal}.
In the pure-state limit all ranks above the ground level are empty,
and the pinned plateaus disappear.

\textit{Numerical evaluation.}
Table~\ref{tab:thermal} is obtained by diagonalizing
Eq.~\Lsector\ in each total-charge sector for
$E_J=\Ec/2$, combining the levels of all sectors
into the Gibbs spectrum, and
evaluating the passive assignments on the two sides of
each crossing.
The ground energies give $\delta_J/\Ec=0.203790$.
The pinned width is the jump of $q_P$ at $\lambda=\Ec$,
and the last column is evaluated from
Eq.~\eqref{eq:dual} with all branches included;
at $\beta\Ec=1$ the charge $q=1/2$ lies outside the
principal interval.
Charge states with $|m_i|\le m_{\max}$ are retained, and
increasing $m_{\max}$ from $8$ to $10$ leaves
all displayed digits unchanged.

\begin{table}[htb]
\caption{
Fixed-spectrum staircase at $E_J=\Ec/2$.
The principal interval is the charge interval
with slope $\lambda=s_0=\Ec-\delta_J
=0.796210\,\Ec$, and
$\Delta q_{\rm pr}=p_1-p_2$ is its width.
The column $\Delta q_{\rm pin}$ gives the total
width of the pinned plateau at $\lambda=\Ec$.
The last column gives the charge-locked ergotropy
at $q=1/2$, in units of $\Ec$.
}
\label{tab:thermal}
\begin{ruledtabular}
\begin{tabular}{ccccc}
$\beta\Ec$
& Principal interval
& $\Delta q_{\rm pr}$
& $\Delta q_{\rm pin}$
& $\W_Q(P,\tfrac12)/\Ec$\\
\hline
$\infty$ & $[0,\,1]$           & $1$      & $0$      & $0.3981$\\
$4$      & $[0.1136,\,0.9889]$ & $0.8753$ & $0.0199$ & $0.3509$\\
$2$      & $[0.3920,\,0.8818]$ & $0.4898$ & $0.1605$ & $0.2309$\\
$1$      & $[0.5465,\,0.7240]$ & $0.1775$ & $0.2764$ & $0.1809$\\
\end{tabular}
\end{ruledtabular}
\end{table}

\textit{Symmetry and constant-current modulation.}
Charge inversion gives $\lambda(-q)=-\lambda(q)$, and the
recurrence~\Lrec\ gives
$\lambda(q+2)=\lambda(q)+2\Ec$ on differentiable branches.
Hence $\lambda(2-q)=2\Ec-\lambda(q)$: the principal interval of slope
$\Ec-\delta_J$ in Table~\ref{tab:thermal} has a partner of slope
$\Ec+\delta_J$, obtained by $q\mapsto2-q$, and all intervals repeat under
$q\mapsto q+2$ with their slopes increased by $2\Ec$.
Along a prescribed discharge $q(t)=q(0)-\gamma t$, the periodic part
$\lambda(q)-\Ec q$ of the spectral slope [Fig.~1(b)]
repeats after $2/\gamma$, at frequency $\gamma/2=I/4e$ with
$I=2e\gamma$.
This counts cycles per unit time and differs from the
angular frequency $|\lambda|/\hbar$ of the phase evolution~\eqref{eq:clock}.
The discharge multiplier approximates the spectral slope only away from
transitions and for a small current-induced offset
(Sec.~\ref{sec:qllg}).

\section{Covariant relaxation and charge control}
\label{sec:qllg}
\supports{``Covariant relaxation'' and ``Discharge'':
the balance~\Lbal, the mobility~\Lmob, the finite-current
offset $\gamma/\mathcal M$, and the loss at a constrained minimum.}

We consider the covariant qLLG equation~\Lqllg\ for the single
charge $Q$, with time-independent $H$ and $Q$.
The multiplier $\lambda(t)$ enters the relaxation law, while the
Hamiltonian $H$ is held fixed, and it is chosen either to
preserve the mean charge or to impose a prescribed
current.

\textit{Spectral solution.}
With the shifted Hamiltonian $K=H-\lambda Q$ and the covariant velocity
$Y=D_t\rho=\dot\rho+(i\lambda/\hbar)[Q,\rho]$,
Eq.~\Lqllg\ reads $i\hbar Y=[K+\hbar\kappa Y,\rho]$.
In the instantaneous eigenbasis $\rho\ket j=p_j\ket j$, write
$A_{jk}=\bra jA\ket k$ and $p_{jk}=p_j-p_k$.
The matrix elements satisfy
$\hbar(i+\kappa p_{jk})Y_{jk}=-p_{jk}K_{jk}$,
so that
\begin{equation}
Y_{jk}
=
\frac{ip_{jk}-\kappa p_{jk}^2}
     {\hbar(1+\kappa^2p_{jk}^2)}\,
K_{jk}.
\label{eq:spectral}
\end{equation}
The matrix $Y$ is Hermitian and vanishes within
each eigenspace of $\rho$.
Equivalently, the laboratory-frame equation is
\begin{equation}
i\hbar\dot\rho=[H+\hbar\kappa Y,\rho].
\label{eq:lab}
\end{equation}
Its generator is Hermitian, so that the evolution
preserves the spectrum of $\rho$, also
for time-dependent $\lambda$.

\textit{Energy balance and charge exchange.}
Taking the trace of Eq.~\Lqllg\ with $Y$ gives
$i\hbar\Tr(Y^2)=\Tr(Y[K,\rho])$, while
Eq.~\eqref{eq:lab} gives the rates of change of the
energy and of the mean charge,
\begin{equation}
\dot E=i\kappa\Tr(Y[H,\rho]),
\qquad
\dot q=\frac{i}{\hbar}\Tr([H,Q]\rho)+i\kappa\Tr(Y[Q,\rho]).
\label{eq:Eqdot}
\end{equation}
Combining these relations,
\begin{equation}
\dot E=\lambda\,i\kappa\Tr(Y[Q,\rho])-\diss,
\qquad
\diss\equiv\kappa\hbar\Tr(Y^2)\ge0.
\label{eq:genbalance}
\end{equation}
For $[H,Q]=0$ the first term of $\dot q$ in
Eq.~\eqref{eq:Eqdot} vanishes, and
Eq.~\eqref{eq:genbalance} is the balance~\Lbal.
The mean charge then changes only through damping, and only if
$[Q,\rho]\ne0$.
The environment responsible for this exchange is not modeled;
$\lambda$ is the charge-transfer potential in the energy balance.
Equations~\eqref{eq:Eqdot} and~\eqref{eq:genbalance} do not use
$[H,Q]=0$ and therefore hold also for the Hamiltonian $H_{\rm prep}$ of
Sec.~\ref{sec:num}, which exchanges charge with the reference; the balance
is then no longer of the form~\Lbal.
Vanishing dissipation implies $Y=0$ and hence
$[H-\lambda Q,\rho]=0$, the condition~\Lstat, which permits
$\dot\rho\ne0$ and is satisfied by every constrained stationary state,
not only by the minimum.

\textit{Current control.}
At each instant, Eq.~\eqref{eq:spectral} is linear in $K$.
We write $Y[A]$ for its right-hand side with $K$ replaced by
$A$, keeping the instantaneous $\rho$ fixed.
Then $Y=Y[H]-\lambda Y[Q]$,
and the mean-charge derivative $\dot q=\Tr(QY)$ is affine in
$\lambda$:
\begin{equation}
\dot q=a+\mathcal M\lambda,
\qquad
a=\Tr(QY[H]),
\qquad
\mathcal M
=-\Tr(QY[Q])
=\frac{\kappa}{\hbar}
\sum_{j,k}
\frac{p_{jk}^2\,|Q_{jk}|^2}
     {1+\kappa^2p_{jk}^2}
\ge0.
\label{eq:mob}
\end{equation}
This is the mobility~\Lmob.
Both $a$ and $\mathcal M$ are evaluated at the
instantaneous state: $a$ is the charge drift at
$\lambda=0$, and $\mathcal M$ is the change
of that drift per unit change of $\lambda$.
Equation~\eqref{eq:mob} does not use $[H,Q]=0$.
When $\mathcal M>0$, the multiplier
$\lambda_0=-a/\mathcal M$ cancels the current, so that
$\dot q=\mathcal M(\lambda-\lambda_0)$ as in the article.
A prescribed discharge rate $\dot q=-\gamma$,
with $\gamma>0$, requires
\begin{equation}
\lambda
=\lambda_0-\frac{\gamma}{\mathcal M}.
\label{eq:lamoffset}
\end{equation}
Thus $\gamma/\mathcal M$ is the finite-current
offset from the instantaneous zero-current value.
For a general state, $\lambda_0$ need not equal the slope of
the constrained minimum.

\textit{Charge-diagonal states.}
For $\kappa>0$, Eq.~\eqref{eq:mob} shows that
$\mathcal M=0$ precisely when $[Q,\rho]=0$.
In this case $Y[Q]=0$ and $a=0$, so that no
finite multiplier can impose a nonzero current, and
$\dot\rho=Y[H]$ independently of $\lambda$.
For $[H,Q]=0$, this evolution preserves all sector weights,
and $a=\mathcal M=0$ throughout the relaxation.
The state nevertheless relaxes within each
sector, with $\diss>0$, whenever $[H,\rho]\ne0$.
The one-sided boost of Sec.~\ref{sec:contact}
is of this type.

\textit{Loss on a minimum branch.}
Consider a constrained minimum $\rho_*$ with
$\mathcal M>0$ in the interior of a linear branch.
Its stationarity condition~\Lstat\ is
$[H-\lambda_0Q,\rho_*]=0$ with
$\lambda_0=\partial_q\emin(P,q)$.
Equation~\eqref{eq:spectral} then gives
$Y[H]=\lambda_0Y[Q]$: the zero-current
multiplier is the spectral slope.
Imposing $\dot q=-\gamma$ produces
$Y=(\gamma/\mathcal M)Y[Q]$, and since
$\kappa\hbar\Tr(Y[Q]^2)=\mathcal M$,
the loss and the balance~\Lbal\ at the minimum are
\begin{equation}
\diss=\frac{\gamma^2}{\mathcal M},
\qquad
-\dot E
=
\left(\lambda_0-\frac{\gamma}{\mathcal M}\right)
\gamma
+\frac{\gamma^2}{\mathcal M}
=\lambda_0\gamma.
\label{eq:lossmin}
\end{equation}
The spectral slope is the open-circuit potential of the article.
The terminal potential is lower by
$\gamma/\mathcal M$, producing the loss rate $\gamma^2/\mathcal M$.
Equation~\eqref{eq:lossmin} holds instantaneously at the minimum.
Away from it, relaxation also contributes and the loss is
$\diss=\kappa\hbar\Tr(Y^2)$.

\textit{Principal branch and its endpoints.}
On an isolated principal branch
(Sec.~\ref{sec:thermal}), the rotated pair~\eqref{eq:ab} is formed from
$\ket{g_M}$ and $\ket{g_{M+1}}$ and carries the probabilities $p_1>p_2$,
while all remaining eigenvectors are common eigenstates
of $H$ and $Q$.
Only the matrix element
$|\bra{\chi_1}Q\ket{\chi_2}|=\sin\vartheta\cos\vartheta$ contributes to the
mobility, and with $q=q_-+(p_1-p_2)\sin^2\vartheta$ Eq.~\eqref{eq:mob}
gives
\begin{equation}
\hbar\mathcal M
=
\frac{2\kappa(q-q_-)(q_+-q)}
     {1+\kappa^2(p_1-p_2)^2},
\qquad
q_+-q_-=p_1-p_2.
\label{eq:parabola}
\end{equation}
The mobility is positive in the branch interior
and vanishes at both endpoints, where the state
becomes charge diagonal.
For ideal tracking of this branch,
$\lambda=s_M-\gamma/\mathcal M$ and $\diss=\gamma^2/\mathcal M$.
For $0<\epsilon<(p_1-p_2)/2$, the loss between
$q_+-\epsilon$ and $q_-+\epsilon$ is
\begin{equation}
L_\epsilon=
\gamma\int_{q_-+\epsilon}^{q_+-\epsilon}\frac{\dd q}{\mathcal M(q)}
=\gamma\hbar\,
\frac{1+\kappa^2(p_1-p_2)^2}
     {\kappa(p_1-p_2)}
\ln\frac{p_1-p_2-\epsilon}{\epsilon}.
\label{eq:ohmic}
\end{equation}
At fixed $\gamma$, the elapsed time $[p_1-p_2-2\epsilon]/\gamma$ stays
finite as $\epsilon\to0$, whereas $L_\epsilon$ diverges
logarithmically and the offset as $1/\epsilon$.
Near an endpoint, $\lambda<0$ means energy input through charge transfer;
no bounded multiplier can sustain the current through that endpoint.
The simulated trajectories need not reach these charge-diagonal states.
Their transition regions and negative excursions are discussed in
Sec.~\ref{sec:num}.

\section{Thermal contact and the two-capacitor benchmark}
\label{sec:contact}
\supports{the bath statement in ``Even charges'', and the
two-capacitor counterpart in ``Locking and the moving minimum''.}

\textit{Thermal contact at fixed charge.}
We consider thermal contact at the reference inverse temperature
$\beta$, through a system--bath interaction that commutes with $Q$.
Since $[H,Q]=0$, the contact preserves the weight
$w_M(\rho)$ of each total-charge sector, and thus the full distribution of
the total charge rather than only its mean.
We further assume complete thermalization within
each sector and decay of the coherences between
different sectors.
With the sector thermal states $\sigma_M$ of
Eq.~\eqref{eq:blocks}, the limiting state is then
\begin{equation}
\mathcal T\rho
=\bigoplus_M w_M(\rho)\,\sigma_M.
\label{eq:T}
\end{equation}
At fixed sector weights, $\mathcal T\rho$ minimizes
the free energy $F(\rho)=E(\rho)-\beta^{-1}S(\rho)$.
Consequently, the average work extracted using
this bath, with controls that preserve the sector
weights and return the Hamiltonian to $H$, obeys
\begin{equation}
W_{\rm ext}
\le F(\rho)-F(\mathcal T\rho).
\label{eq:DeltaF}
\end{equation}
The joint shift $B$ leaves the relative Hamiltonian
unchanged, so that $B^k\sigma_{M-2k}B^{-k}=\sigma_M$, and the boosted
Gibbs state at mean charge $q=2k$ is
\begin{equation}
\rho_*
=B^k\rho_GB^{-k}
=\bigoplus_M w_{M-2k}^G\,\sigma_M.
\label{eq:boosted}
\end{equation}
It is already thermal within every charge sector, so that
$\mathcal T\rho_*=\rho_*$ and the bound~\eqref{eq:DeltaF}
vanishes.
No work can be extracted from this state
using the bath under these restrictions, and the charge-locked energy
$2\Ec k^2$ remains stored, as stated in the article.
At other mean charges, minimizers can carry intersector coherence.
The assumed contact removes it and generally changes the spectrum.

\textit{Two-capacitor counterpart.}
The article shows that the one-sided boost
$\rho_0=T_1^2\rho_GT_1^{-2}$ and the balanced minimum
$\rho_*=B\rho_GB^\dagger$ at $q=2$ have
$E(\rho_0)-E_G=4\Ec$ and $E(\rho_*)-E_G=2\Ec$
for every $\beta>0$ and $E_J\ge0$, and that the charge-conserving unitary
$U=T_1^\dagger T_2$ maps $\rho_0$ to $\rho_*$.
In the classical counterpart, two identical capacitors of capacitance
$C_\Sigma$ initially carry the charges $4e$ and zero.
Their electrostatic energy $(4e)^2/(2C_\Sigma)=4\Ec$ falls to
$2(2e)^2/(2C_\Sigma)=2\Ec$ after equal redistribution of the charge, and
in a relaxing circuit the difference is
transferred to other degrees of freedom, for example
as Joule heat or radiation~\cite{Zucker1955,Mita1999,Boykin2002}.
The quantum preparation has the same partition of the added
energy: half is extractable at fixed mean charge and half is
charge-locked, for every reference temperature and Josephson coupling.

\textit{Release of the free half.}
Under the qLLG flow, $\rho_0$ is a charge-diagonal state
(Sec.~\ref{sec:qllg}): the multiplier drops out, and by
Eq.~\Lbal\ the loss integrated at fixed $q=2$ is
$2\Ec-[E(t)-E(\rho_*)]$, which approaches $2\Ec$ only if the trajectory
reaches the balanced minimum (Sec.~\ref{sec:num}).
The thermal contact~\eqref{eq:T} does reach it:
the sector weights of $\rho_0$ are $w_M(\rho_0)=w_{M-2}^G$, so that
$\mathcal T\rho_0=\rho_*$ by Eq.~\eqref{eq:boosted}, and with the
Hamiltonian held fixed and no work extracted the bath receives the heat
$2\Ec$.
Alternatively, the free half is extracted as the work
$E(\rho_0)-E(\rho_*)=2\Ec$ by any cyclic control $H+V(t)$, with
$[V(t),Q]=0$, that implements $U$.
In either case, $2\Ec$ remains charge-locked in
$\rho_*$ for the fixed Hamiltonian $H$.
At $E_J=0$ the partition is unchanged, but $[H,\rho_0]=0$ and
the qLLG flow does not relax $\rho_0$.
The capacitively coupled absorber of Sec.~\ref{sec:circuit} cannot
relax $\rho_0$ either,
because its coupling then preserves each island charge.

\section{Numerical integrations}
\label{sec:num}
\supports{the preparations, phase rates, plateau
offsets, and energy partition in ``Locking and the
moving minimum'' and ``Discharge'' [Fig.~2].}

\textit{Numerical evolution.}
We represent the density matrix in the two-island
charge basis $\ket{m_1,m_2}$, retaining states with
$|m_i|\le m_{\max}$, and use $E_J=\Ec/2$ and $\kappa=1/2$
throughout.
At each evaluation of the evolution equation,
we diagonalize $\rho$, obtain the covariant
velocity $Y$ from Eq.~\eqref{eq:spectral}, and set
$\dot\rho=Y-(i\lambda/\hbar)[Q,\rho]$.
By Eq.~\eqref{eq:mob} the multiplier is $\lambda=-a/\mathcal M$ for
$\dot q=0$ and $\lambda=-(a+\gamma)/\mathcal M$ for $\dot q=-\gamma$, with
$a$ and $\mathcal M$ evaluated
at the instantaneous state rather than on the constrained minimum.
When $\hbar\mathcal M$ falls below a round-off threshold,
$10^{-14}$ in general and $10^{-12}$ during the charging
stage described below, $\lambda$ is set to zero.
We integrate with the explicit Runge--Kutta method
DOP853~\cite{Hairer1993,SciPy2020}, using relative and
absolute tolerances $(10^{-8},10^{-10})$ for the preparation
and the fixed-charge relaxation, and $(10^{-7},10^{-9})$ for
the discharge.
The inverse temperature $\beta$ enters only through the
spectrum $P$ of the initial Gibbs state, which the flow preserves.

\textit{Preparations.}
All runs start from the one-sided boost
$\rho_0=T_1^2\rho_GT_1^{-2}$ of the Gibbs state, which has $q=2$ and
commutes with $Q$.
For the fixed-charge run at $q=2$ [Fig.~2(a,b)], $\rho_0$ is
the initial state.
By Sec.~\ref{sec:qllg}, $a$ and $\mathcal M$ vanish
for this state and at all later times, and the multiplier
drops out of the evolution.
Numerically, round-off generates coherences between sectors
below $3\times10^{-10}$, which do not affect the trajectory, and no
multiplier is shown for this run.

To raise $q$ from $2$ to $2.5$ with the qLLG flow, we create
intersector coherence in a charging stage that couples island~1 to the
phase reference of Sec.~\ref{sec:ref}:
\begin{equation}
H_{\rm prep}
=H-E_x\cos\varphi_1
=H-\frac{E_x}{2}(T_1+T_1^\dagger),
\qquad E_x=0.3\,\Ec.
\label{eq:Hprep}
\end{equation}
Starting from $\rho_0$, we integrate Eq.~\Lqllg\ with
$H_{\rm prep}$ and the same $\kappa$, choosing the multiplier from
Eq.~\eqref{eq:mob} to impose $\dot q=\Ec/(150\hbar)$ until
$q=2.5$, at approximately $75\,\hbar/\Ec$.
Initially $\mathcal M=0$, so the first evaluations use
$\lambda=0$; the multiplier is large while coherence develops.
Removing the contact defines $t=0$ for the fixed-charge run, which then
uses $H$.
The prepared state remains on the Gibbs orbit $\Op$.
For each discharge run [Fig.~2(c,d)], it is first relaxed at
$q=2.5$ for $300\,\hbar/\Ec$, separately at each temperature; the
discharge clock starts after this relaxation.

\textit{Observables.}
To measure the approach to the constrained
minimum, we define the excess energy
\begin{equation}
X(t)=E(\rho(t))-\emin(P,q(t))=\W_{\rm free}(\rho(t)).
\label{eq:X}
\end{equation}
This is the first term of Eq.~\Lsplit\ and the quantity $E-\emin$
in Fig.~2(a).
For the run at $q=2$, Fig.~2(a) also shows the charge
imbalance $\Delta N=\langle N_1\rangle-\langle N_2\rangle
=2\langle n\rangle$ and the scaled Josephson signal
$4\langle\sin\theta\rangle=4\operatorname{Im}\langle T_1T_2^\dagger\rangle$,
which is proportional to the pair current
between the islands.
The rate of the relative phase in Fig.~2(b) is the time
derivative of
$\arg\langle e^{i\theta}\rangle=\arg\langle T_1T_2^\dagger\rangle$.
The charging term gives its ac-Josephson estimate
$4\Ec\langle n\rangle/\hbar=2\Ec\Delta N/\hbar$.
For $q=2.5$, the common-phase rate is the derivative of
$\arg\langle T_1\rangle$; at the moving minimum it equals
$\lambda/\hbar$ by Eq.~\eqref{eq:clock}.
The rates are obtained by differentiating the unwrapped
phases on the output grid.

\textit{Fixed-charge relaxation.}
The two runs of Fig.~2(a,b) use $\beta\Ec=2$ and
$m_{\max}=5$, and extend to $t=300\,\hbar/\Ec$; the figure
shows the first third.
For $q=2$, the excess energy decreases from
$X(0)=2\Ec$, the free ergotropy of the one-sided boost, to $0.088\,\Ec$ at
$t=100\,\hbar/\Ec$ and $0.020\,\Ec$ at $t=300\,\hbar/\Ec$,
while $\Delta N$ decays from $2$ to $0.11$ and $0.02$.
For $q=2.5$, $X$ decreases from $0.290\,\Ec$ to
$0.098\,\Ec$ and $0.033\,\Ec$ at the same times.
The constrained minimum of this run lies on the principal branch
$2.392\le q\le2.882$ (Table~\ref{tab:thermal}), with slope
$s_2=3\Ec-\delta_J=2.7962\,\Ec$.
Over the plotted interval $0\le t\Ec/\hbar\le100$,
including the initial transient, the mean rate of
$\arg\langle T_1\rangle$ is $2.788\,\Ec/\hbar$,
$0.3\%$ below $s_2/\hbar$, as quoted in the article.
Over $60\le t\Ec/\hbar\le300$, the mean rates of
$\arg\langle T_1\rangle$ and
$\arg\langle T_2\rangle$ are
$2.7960\,\Ec/\hbar$ and $2.7969\,\Ec/\hbar$,
within $0.03\%$ of $s_2/\hbar$.
Instantaneous rates fluctuate by up to $1.4\%$, and the mean
multiplier is $2.7949\,\Ec$.
Over the same interval $|\langle T_1T_2^\dagger\rangle|$
stays between $0.25$ and $0.27$, and its phase within
$\pm0.05$~rad, consistent with the stationarity of
the relative coherence at the minimum (Sec.~\ref{sec:ref}).
The remaining positive values of $X$ show that
neither run has reached its constrained
minimum by the final integration time.

\textit{Finite-current offsets.}
The discharge runs impose the constant rate
$\dot q=-\gamma$ with
$\gamma=2.5\times10^{-4}\,\Ec/\hbar$, taking the mean
charge from $q=2.5$ to $q=0.05$ in the time
$9800\,\hbar/\Ec$; they use $m_{\max}=4$.
The three principal plateaus encountered along
this path have the slopes $s_2$, $s_1$, and $s_0$ of
Eq.~\eqref{eq:slopes}.
By Eq.~\eqref{eq:lamoffset}, the multiplier at a state with
$\mathcal M>0$ is $\lambda=\lambda_0-\gamma/\mathcal M$,
and $\lambda_0=s_M$ only on an exact principal minimum.
Thus $s_M-\lambda=\gamma/\mathcal M+(s_M-\lambda_0)$.
Since the instantaneous mobility was not stored, we compare the mean
deficit $s_M-\overline{\lambda}$ with a branch estimate
$\overline{\gamma/\mathcal M_{\rm br}}$.
Here $\mathcal M_{\rm br}(q)$ is Eq.~\eqref{eq:parabola} at the recorded
charge, and bars denote time averages over the same samples.
The averages cover the central half of each principal
interval, for which Eq.~\eqref{eq:parabola} gives
\begin{equation}
\overline{\frac{\gamma}{\mathcal M_{\rm br}}}
=2\ln3\;
\frac{\gamma\hbar\,[1+\kappa^2(p_1-p_2)^2]}
     {\kappa\,(p_1-p_2)^2}.
\label{eq:centralhalf}
\end{equation}
On the first plateau, only the portion
$2.33\le q\le2.5$ is reached by the discharge,
and the averages are restricted accordingly.
At $\beta\Ec=4$ [Fig.~2(c)] the deficits on the plateaus $s_2$, $s_1$,
and $s_0$ are $1.67$, $1.58$, and $1.71$, against the branch estimates
$1.75$, $1.71$, and $1.71$, all in units of $10^{-3}\Ec$; they agree
to within $8\%$.
At $\beta\Ec=2$ the plateaus are narrower,
$p_1-p_2=0.49$, and the trajectory is farther from the
minimum.
The two fully traversed principal plateaus then have the
deficits $4.0\times10^{-3}\Ec$ and
$1.7\times10^{-3}\Ec$, against the branch
estimate $4.9\times10^{-3}\Ec$ for each, an agreement only within a
factor of three.

\textit{Energy accounting.}
For a discharge at constant $\gamma>0$, we write
$\Delta E_{\rm rel}(t)=E(0)-E(t)$ for the released energy, denoted
$E_0-E(t)$ in the article, and
$R_Q(t)=\gamma\int_0^t\lambda(t')\,\dd t'$ and
$L(t)=\int_0^t\diss(t')\,\dd t'$ for its signed charge-transfer
contribution and its internal loss.
The balance~\Lbal\ then gives
$\Delta E_{\rm rel}(t)=R_Q(t)+L(t)$, the decomposition shown
in Fig.~2(d).
With the spectral slope $\lambda_{\min}(q)=\partial_q\emin(P,q)$, the
excess energy~\eqref{eq:X} satisfies
$\dot X=\gamma[\lambda_{\min}(q)-\lambda]-\diss$.
The area between the spectral staircase and
the dynamical multiplier in Fig.~2(c) thus equals the
accumulated loss, up to the change of the excess energy.
Table~\ref{tab:loss} lists the totals for the two runs of
Fig.~2(c); Fig.~2(d) shows the run at $\beta\Ec=4$.
The sampled loss fractions $0.080$ and $0.159$ give the approximate $8\%$
and $16\%$ quoted in the article; the resolution is discussed below.

\begin{table}[htb]
\caption{
Energy accounting for the prescribed-current
discharges of Fig.~2(c,d), from $q=2.5$ to $q=0.05$ at
$\gamma=2.5\times10^{-4}\,\Ec/\hbar$.
$X(0)$ is the excess energy~\eqref{eq:X} of the
initial state and $X_{\max}$ its maximum along the run.
The final column is the fraction of the released
energy assigned to internal loss.
}
\label{tab:loss}
\begin{ruledtabular}
\begin{tabular}{ccccccc}
$\beta\Ec$
& $X(0)/\Ec$
& $X_{\max}/\Ec$
& $\Delta E_{\rm rel}/\Ec$
& $R_Q/\Ec$
& $L/\Ec$
& $L/\Delta E_{\rm rel}$\\
\hline
$4$ & $0.030$ & $0.085$
& $3.284$ & $3.019$ & $0.263$ & $0.080$\\
$2$ & $0.024$ & $0.051$
& $3.237$ & $2.726$ & $0.516$ & $0.159$\\
\end{tabular}
\end{ruledtabular}
\end{table}

\textit{Resolution and checks.}
The multiplier becomes negative in parts of the transition regions.
About $90\%$ of $L$ accumulates outside the central halves of the
principal intervals.

We evaluate $R_Q$ and $L$ by trapezoidal
quadrature on the output grid of 2001 points, whose
spacing $4.9\,\hbar/\Ec$ resolves the narrowest transition
regions with only a few samples.
The residual $R_Q+L-\Delta E_{\rm rel}$ is approximately
$-2\times10^{-3}\Ec$ at $\beta\Ec=4$ and
$4\times10^{-3}\Ec$ at $\beta\Ec=2$.
Cancellation in this sum does not establish the separate accuracy of $R_Q$
and $L$.
For $0\le q\le2$, the static recurrence~\Lrec\ has truncation errors
below $6\times10^{-9}\Ec$ at $m_{\max}=5$, $\beta\Ec=2$, and of order
$10^{-6}\Ec$ and $10^{-5}\Ec$ at $m_{\max}=4$, $\beta\Ec=4$ and $2$,
respectively.
The preparation is more sensitive to the cutoff: at $\beta\Ec=2$ the
state relaxed at $q=2.5$ for $300\,\hbar/\Ec$ has $X=0.033\,\Ec$ for
$m_{\max}=5$ and $0.024\,\Ec$ for $m_{\max}=4$ (Table~\ref{tab:loss}).
In the fixed-charge runs the mean charge drifts by less than
$10^{-12}$ and the density-matrix spectrum by less than
$10^{-6}$, and in the discharge runs the mean charge follows
$q(0)-\gamma t$ to within $10^{-13}$.
The code, the saved trajectories, and the
diagnostic scripts are available in
Ref.~\cite{code}.

\section{Outline of a circuit realization}
\label{sec:circuit}
\supports{the preparations and phase frequencies in ``Locking and the
moving minimum'', and the circuit outline mentioned in ``Summary''.}

\begin{figure}[tp]
\centering
\includegraphics[width=0.94\linewidth]{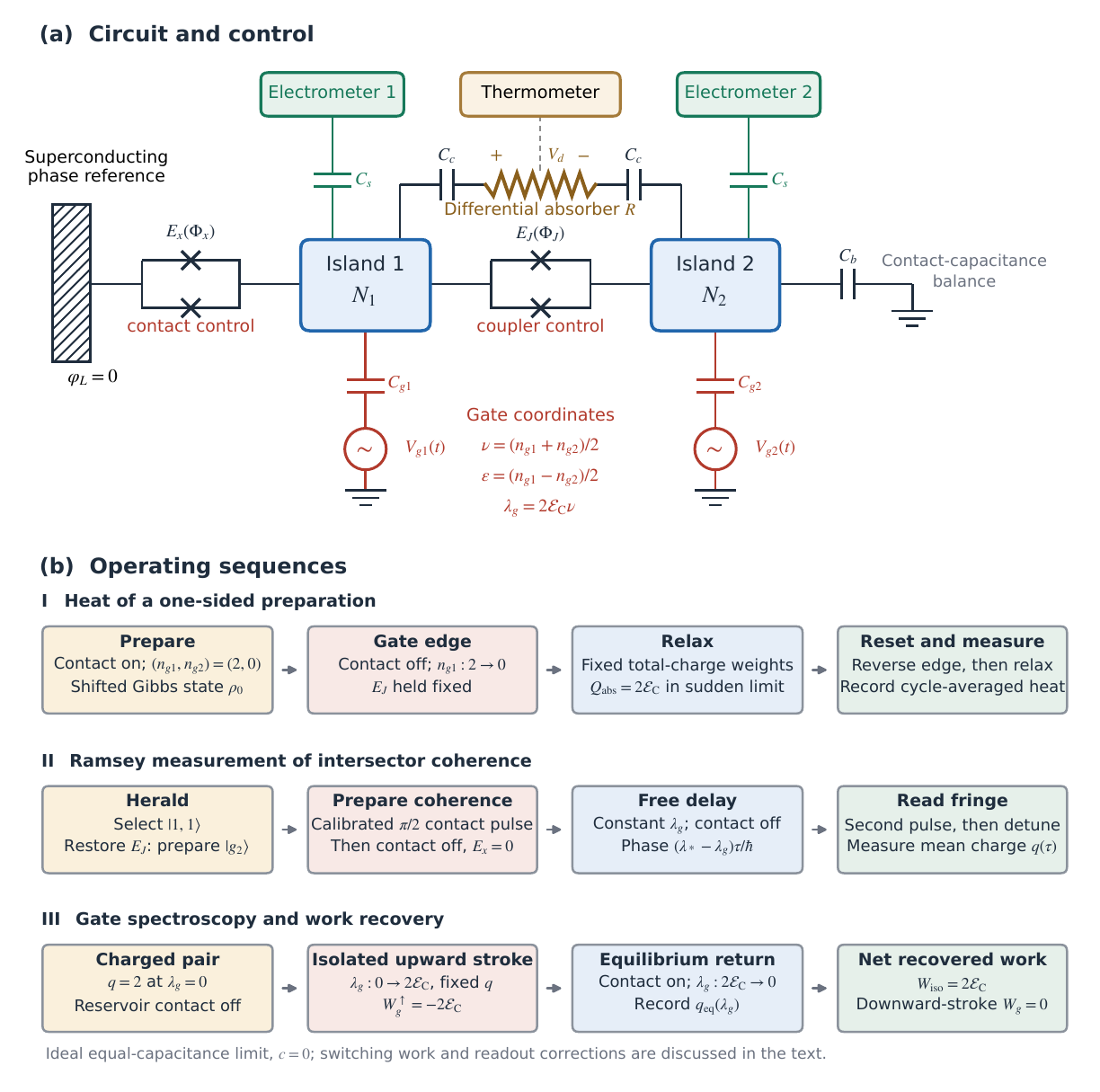}
\caption{(a) Circuit and controls. The contact $E_x$ exchanges
Cooper pairs with the phase reference; $E_J$ transfers them between islands.
The absorber couples through $C_c$ to the differential voltage;
$C_b$ balances the contact capacitance. Red and green denote control and readout.
(b) The three sequences of Sec.~\ref{sec:circuit}. The Ramsey phase
increment is that of $\langle T_1\rangle$, opposite to $\Delta\zeta$.}
\label{fig:circuit}
\end{figure}

Figure~\ref{fig:circuit} outlines a circuit and three sequences
that test the energy
and spectral relations of the article.
The reservoir contact and electromagnetic environment set the dynamics;
these sequences do not implement the isospectral qLLG relaxation law.
We use $E_J=\Ec/2$ and $\Ec/h=2.50$~GHz ($C_\Sigma=31.0$~fF), for which
$\beta\Ec=4$ and $2$ correspond to $30$ and $60$~mK.
The energy balances below assume weak contacts and complete relaxation;
contact-switching and interaction energies require separate accounting.

\textit{Circuit.}
Two superconducting islands are coupled by a SQUID with Josephson energy
$E_J(\Phi_J)$, controlled by the flux $\Phi_J$.
A second SQUID, with energy
$E_x(\Phi_x)$ controlled by $\Phi_x$, connects island~1 to the Cooper-pair
reservoir and phase reference
$\varphi_L=0$ of Sec.~\ref{sec:ref}.
The contact is open at $E_x=0$; $C_b$ balances its capacitance on
island~2.
Gate voltages $V_{gi}(t)$ through $C_{gi}$ set the
offset charges $n_{gi}$, in Cooper-pair
units~\cite{Nakamura1999,Pashkin2003}.
Electrometers coupled through $C_s$ measure the island
charges~\cite{Schoelkopf1998}.
A normal-metal strip of resistance $R$, coupled through equal capacitors
$C_c$ and monitored by a thermometer, absorbs energy.
For identical islands with negligible mutual capacitance,
\begin{equation}
H_{\rm circ}
=H-\lambda_g Q-4\Ec\varepsilon n
+\frac{\lambda_g^2}{2\Ec}+2\Ec\varepsilon^2
-E_x\cos\varphi_1+H_{\rm int}+H_R,
\qquad
\lambda_g=2\Ec\nu,
\label{eq:Hgate}
\end{equation}
where $\nu=(n_{g1}+n_{g2})/2$ and
$\varepsilon=(n_{g1}-n_{g2})/2$.
The scalar gate terms are retained for the work balance.
Here $H_R$ describes the thermal absorber, and
$H_{\rm int}=-2e\eta nV_d$ couples the relative charge to the strip
voltage $V_d$, with $\eta=C_c/C_\Sigma$.
Thus $\lambda_g$ enters as $\lambda$ does in $H-\lambda Q$, but is a
physical gate control.
With the contact open, $[Q,H_{\rm circ}]=0$ fixes the sector weights.
In the weak-contact limit after equilibration, $\nu=k$, $\varepsilon=0$
prepares $B^k\rho_GB^{-k}$; the setting
$(n_{g1},n_{g2})=(2,0)$ prepares $\rho_0=T_1^2\rho_GT_1^{-2}$, the Gibbs
state of $T_1^2HT_1^{-2}$.

\textit{Sequence I: heat of a one-sided preparation.}
After equilibration at $(n_{g1},n_{g2})=(2,0)$, the contact is
opened and $n_{g1}$ is stepped to zero with $E_J$ held fixed.
An ideal sudden edge leaves $\rho_0$ unchanged, restores $H$, and supplies
$4\Ec$ of gate work.
Complete within-sector relaxation at the reference temperature gives
$\mathcal T\rho_0=\rho_*$ (Sec.~\ref{sec:contact}), releasing
$Q_{\rm abs}=\W_{\rm free}(\rho_0)=2\Ec$ as heat while $2\Ec$
remains locked.
The reverse edge changes the Hamiltonian back and does no work;
subsequent relaxation releases another $2\Ec$.
The full cycle therefore supplies and dissipates $4\Ec$.
Repeated cycles provide a mean-power signal for
bolometry~\cite{Kokkoniemi2019}.
Finite gate edges require including the unitary evolution during the edge;
at $E_J=0$, this absorber cannot relax the charge preparation.

\textit{Sequence II: Ramsey measurement of intersector coherence.}
At $\nu=1$, the contact is opened and the inter-island coupling
suppressed.
Selecting $(N_1,N_2)=(1,1)$ by charge readout, then increasing $E_J$
adiabatically, prepares the pure sector ground state $\ket{g_2}$.
This conditioning changes the thermal density-matrix spectrum.
A weak contact couples $\ket{g_2}$ and $\ket{g_3}$ through
$-E_xu/2$, with $u=\bra{g_3}T_1\ket{g_2}$ and $|u|=0.785$.
At $\lambda_g=\lambda_*\equiv s_2=3\Ec-\delta_J$, these
states are degenerate in $H-\lambda_gQ$.
A calibrated contact pulse near this crossing prepares their equal
superposition, the pure-state minimum~\eqref{eq:purestate} at $q=2.5$.
The contact is opened for a delay $\tau$.
A second pulse converts the phase $\zeta$ of this superposition into a charge
fringe at
angular frequency $|\lambda_*-\lambda_g|/\hbar$.
Varying the common gate therefore determines
$E_0(3)-E_0(2)$ and $\delta_J$.
For $E_x=0.1\Ec$, the dressed gap $h\times194$~MHz suggests a
$1.3$~ns $\pi/2$ pulse.

\textit{Sequence III: gate spectroscopy and work recovery.}
With a weak contact and sufficient equilibration time, the common
gate probes $q_{\rm eq}(\lambda_g)$ of
Sec.~\ref{sec:thermal}.
Because the thermal probabilities then vary with the gate, this
scan does not measure the
fixed-spectrum staircase $q_P(\lambda)$.
The same controls permit ideal work recovery.
At $\varepsilon=0$, Eq.~\eqref{eq:Hgate} gives
$\dd W_g=(-q+\lambda_g/\Ec)\,\dd\lambda_g$, with $W_g>0$ for work supplied
to the pair.
Starting from the boosted Gibbs state at $q=2$ and zero gate,
$\lambda_g$ is raised to $2\Ec$ with the contact open, giving
$W_g^{\uparrow}=-2\Ec$, then lowered reversibly with
charge exchange enabled.
The symmetry
$q_{\rm eq}(2\Ec-\lambda_g)=2-q_{\rm eq}(\lambda_g)$ gives
$W_g^{\downarrow}=0$.
The two strokes return
$W_{\rm iso}=2\Ec=\W_Q(P,2)$ to the gate
source, restore zero gate, and discharge the pair by charge exchange.

\textit{Device requirements.}
For identical islands, a mutual capacitance $C_m=cC_\Sigma$ changes the
relative charging term to $2\Ec n^2/(1+2c)$; here $C_\Sigma$ excludes
$C_m$.
The recurrence~\Lrec, even-charge values~\Leven, and
bounds~\Lbounds\ remain exact, while the relative spectrum
changes and the one-sided boost locks the
fraction $(1+2c)/(2+2c)$ of the added energy.
Capacitance mismatch breaks the exact recurrence.
Charge readout requires suppressing or detuning residual tunneling.
The capacitance matrix, residual couplings, gate noise,
quasiparticle parity lifetime~\cite{Aumentado2004}, and readout
backaction must be calibrated for the device.